\documentclass[fleqn,usenatbib]{mnras}
\usepackage{anyfontsize}
\usepackage[T1]{fontenc}
\DeclareRobustCommand{\VAN}[3]{#2}
\let\VANthebibliography\thebibliography
\def\thebibliography{\DeclareRobustCommand{\VAN}[3]{##3}\VANthebibliography}
\usepackage{graphicx}
\usepackage{amsmath}
\usepackage{amssymb}
\usepackage{newtxtext,newtxmath}

\defcitealias{2022MNRAS.511.2948M}{MS22}

\title[Quenching of EAGLE satellites]{Star formation and stellar \& AGN feedback in the absence of accretion, not gas stripping, set the quenching timescale in satellite galaxies}

\author[A. I. Visser-Zadvornyi et al.]{
Anatolii I. Visser-Zadvornyi,$^{1}$
Mary E. Carstairs$^{2,3}$,
Kyle A. Oman$^{1,2,3}$\thanks{kyle.a.oman@durham.ac.uk}
\& Marc A. W. Verheijen$^{1}$
\\
$^{1}$Kapteyn Astronomical Institute, University of Groningen, Postbus 800, NL-9700 AV Groningen, the Netherlands\\
$^{2}$Institute for Computational Cosmology, Physics Department, Durham University, South Road, Durham DH1 3LE, United Kingdom\\
$^{3}$Centre for Extragalactic Astronomy, Physics Department, Durham University, South Road, Durham DH1 3LE, United Kingdom\\
}

\date{Accepted XXX. Received YYY; in original form ZZZ}

\pubyear{2024}

\begin{document}
\label{firstpage}
\pagerange{\pageref{firstpage}--\pageref{lastpage}}
\maketitle

\begin{abstract}
Observational measurements hint at a peak in the quenching timescale of satellite galaxies in groups and clusters as a function of their stellar masses at $M_{\star} \approx 10^{9.5} \mathrm{M}_{\odot}$; less and more massive satellite galaxies quench faster. We investigate the origin of these trends using the EAGLE simulation in which they are qualitatively reproduced for satellites with $10^{9}<M_{\star}/\mathrm{M}_\odot<10^{11}$ around hosts of $10^{13}<M_\mathrm{200c}/\mathrm{M}_\odot<10^{14.6}$. We select gas particles of simulated galaxies at the time that they become satellites and track their evolution. Interpreting these data yields insights into the prevailing mechanism that leads to the depletion of the interstellar medium (ISM) and the cessation of star formation. We find that for satellites across our entire range in stellar mass the quenching timescale is to leading order set by the depletion of the ISM by star formation and stellar \& AGN feedback in the absence of sustained accretion of fresh gas. The turnover in the quenching timescale as a function of stellar mass is a direct consequence of the maximum in the star formation efficiency (or equivalently the minimum in the total -- stellar plus AGN -- feedback efficiency) at the same stellar mass. We can discern the direct stripping of the ISM by ram pressure and/or tides in the simulations; these mechanisms modulate the quenching timescale but do not drive its overall scaling with satellite stellar mass. Our findings argue against a scenario in which the turnover in the quenching timescale is a consequence of the competing influences of gas stripping and `starvation'.
\end{abstract}


\section{Introduction}\label{sec:introduction}

Satellite galaxies of massive groups and clusters are preferentially quenched compared to similar galaxies in the field \citep[e.g.][amongst many others]{1999ApJ...527...54B,2006MNRAS.373..469B,2010ApJ...721..193P,2012MNRAS.424..232W,2020MNRAS.499..230B}. Some consensus has built up around a `delayed then rapid' quenching scenario for such satellites \citep{2000ApJ...540..113B,2010MNRAS.404.1231V,2010MNRAS.406.2249W,2012MNRAS.423.1277D,2013MNRAS.432..336W,2014MNRAS.444.2938H,2015ApJ...806..101H,2016MNRAS.463.3083O,2019A&A...621A.131M,2020ApJS..247...45R,2021MNRAS.501.5073O,2023MNRAS.522.1779R,2023MNRAS.523.6020W}, although it is often difficult to discriminate between a fast transition in star formation rate and a slower transition that sees galaxies within a population cross a given threshold with small scatter \citep{2021MNRAS.501.5073O}. This complicates the interpretation of the `delayed then rapid' scenario because a fast transition is tempting to qualitatively associate with a process strongly stripping the ISM near pericentric passage, while a slower transition that is coherent across a population evokes starvation-driven quenching. This ambiguity is difficult to resolve observationally \citep{2021PASA...38...35C}. Statistical population-based studies generally find that models of both kinds provide acceptable descriptions of the data \citep[but see][]{2019A&A...621A.131M,2023MNRAS.522.1779R}, while detailed studies of individual heavily-stripped objects are difficult to generalise to the broader satellite galaxy population.

\begin{figure*}
  \includegraphics[width=0.9\textwidth]{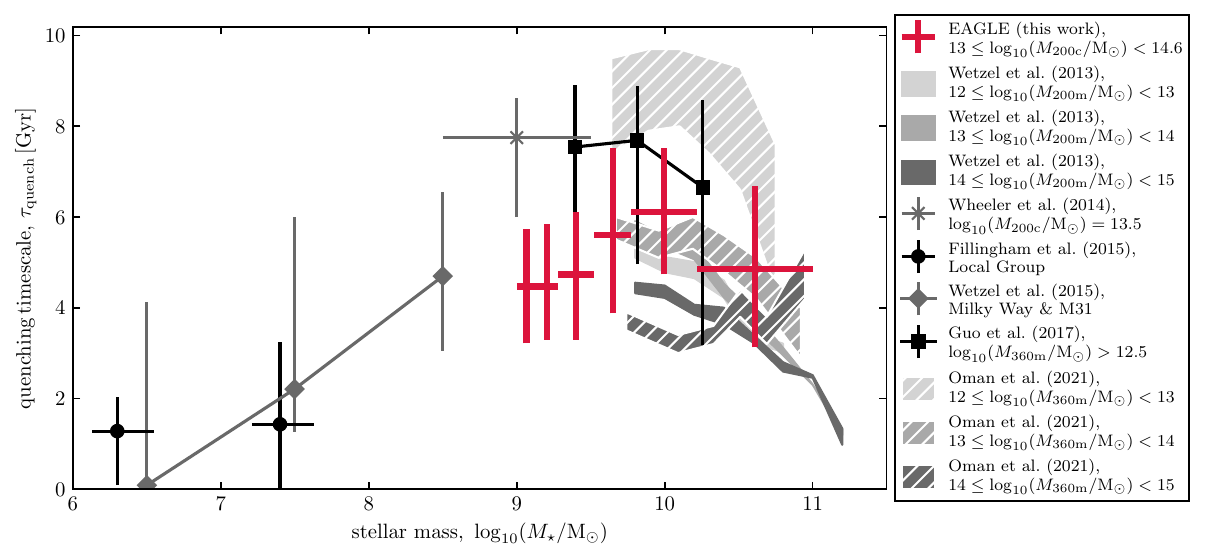}
  \caption{Quenching timescale measurements as a function of satellite stellar mass. Symbols in grey and black show observational results from \citet{2013MNRAS.432..336W,2014MNRAS.442.1396W,2015MNRAS.454.2039F,2015ApJ...808L..27W,2017ApJ...841L..22G,2021MNRAS.501.5073O}, as indicated in the legend. Shaded bands and error bars represent the measurement uncertainties as reported by the cited work. We stress that the definitions and methodologies involved in the various studies are heterogenous. In particular, the timescale is measured for satellites around a variety of types of host systems, as noted in the legend. No effort has been made to correct for systematic differences between different data sources. We highlight the qualitative trend that there seems to be a positive trend at low stellar mass that turns over into a negative trend at high stellar mass. The red crosses mark the median quenching timescale in the intervals of stellar mass shown by the horizontal bars with the interquartile scatter shown by the vertical bars (the uncertainty in the median is too small to show on the figure) for our sample of satellite galaxies drawn from the EAGLE simulations that reproduce the same qualitative trend.}
    \label{fig:obs-qtd}
\end{figure*}

A promising hint from statistical analyses of galaxy populations, and the central focus of this work, is the dependence of the quenching timescale on satellite stellar mass. `Quenching timescale' has been variously defined in many studies. In this work we define it as the time interval between when a galaxy becomes a satellite of its current ($z=0$) host galaxy and when it ceases active star formation. The more specific definitions of these events also vary from study to study. We set these differences aside for the purpose of this introduction to focus on a qualitative feature \citep[but see][sec.~6.2 for a careful comparison accounting for differences in definition]{2021MNRAS.501.5073O}. Fig.~\ref{fig:obs-qtd} shows a compilation of measurements \citep[greyscale markers and regions;][see also \citealp{2024MNRAS.527.9715R} for a complementary measurement of what is probably the same underlying trend]{2013MNRAS.432..336W,2014MNRAS.442.1396W,2015MNRAS.454.2039F,2015ApJ...808L..27W,2017ApJ...841L..22G,2021MNRAS.501.5073O} of the quenching timescale $\tau_\mathrm{quench}$ as a function of satellite stellar mass, spanning a wide range in host system mass.

The feature that we highlight in Fig.~\ref{fig:obs-qtd} is that the quenching timescale has a positive slope as a function of stellar mass at $M_\star\lesssim 10^{9.5}\,\mathrm{M}_\odot$ that turns over to a negative slope at higher stellar masses (see also \citealp{2015MNRAS.454.2039F}, fig.~6, and \citealp{2017ApJ...841L..22G}, fig.~5). This compilation focuses on studies of the local Universe ($z\sim 0$), but similar qualitative trends may hold out to at least $z\sim1$ \citep{2017ApJ...841L..22G}. This could signal a transition between two dominant physical processes setting the quenching timescale. Indeed, it has been used to argue that quenching in low-mass satellites is regulated by ram-pressure or other stripping processes while in massive satellites starvation is dominant \citep{2015MNRAS.454.2039F,2016MNRAS.463.1916F,2017ApJ...841L..22G,2020MNRAS.499..230B,2019MNRAS.487.3740W}. However, the location of the transition is also the stellar mass scale where the stellar-to-halo mass ratio peaks \citep[e.g.][]{2013MNRAS.428.3121M}, thought to be related to a transition between the dominance of stellar and AGN feedback processes \citep[e.g.][hereafter \citetalias{2022MNRAS.511.2948M}, and references therein]{2022MNRAS.511.2948M}. Feedback is thought to play an important role in the quenching of central (i.e. non-satellite) galaxies \citep{2006MNRAS.365...11C,2006MNRAS.366..499D,2008ApJS..175..356H}, so the idea that it could be equally important in satellites is well-motivated \citep{2014MNRAS.442L.105M}.

The review of \citet{2021PASA...38...35C} singles out the fate of gas in the central regions of satellite galaxy discs (that least susceptible to direct stripping) after the first pericentric passage through the host system as one of the most important unknowns in our understanding of environmental quenching. They propose that one possible, perhaps even likely, fate of this gas is to be expelled or heated by feedback. There is some observational evidence both for \citep{2014MNRAS.442L.105M,2018A&A...620A.164B} and against \citep{2017MNRAS.466.1275B,2008MNRAS.387...79V,2019MNRAS.483.5444D} feedback-driven quenching in individual satellite galaxies.

In this work we take the approach of turning to cosmological hydrodynamical simulations to inform our interpretation. The simulations that we use lend strong support to feedback-driven quenching of satellite galaxies at the population level. While direct stripping of the ISM clearly plays a role in the quenching process and may for occasional satellites be the driving mechanism in terms of setting the quenching timescale, we will show that it is primarily the loss of gas through feedback from star formation and/or AGN activity that regulates the satellite quenching timescale (given that they are prevented from accreting fresh gas by their local environment). Our analysis is largely detached from observational constraints which only enter indirectly through their influence on the calibration of the simulations. This naturally means that our conclusions are only as reliable as the simulations that they are based on, but we will argue below that accounting for some of the shortcoming of the simulations that we use would likely strengthen our conclusions.

This article is organised as follows. In Sec.~\ref{sec:methods} we introduce the simulations that we use for our analysis, our methods for tracking the gas particles belonging to satellite (and central) galaxies, and an analytic model that we use to guide our interpretation. In Sec.~\ref{sec:results} we show that quenching of satellites in the simulations is predominantly regulated by feedback, but also explore the secondary role played by gas stripping processes. We discuss caveats in Sec.~\ref{sec:discussion} and summarize in Sec.~\ref{sec:conclusions}.

\section{Methods}\label{sec:methods}

The EAGLE\footnote{Evolution and Assembly of GaLaxies and their Environments.} simulations \citep{2015MNRAS.446..521S} are a suite of cosmological hydrodynamical simulations of galaxy formation and evolution. The simulations adopt the pressure-entropy formulation of smoothed-particle hydrodynamics \citep{2013MNRAS.428.2840H} with a modified time-stepping criterion \citep{2012MNRAS.419..465D} and artificial viscosity \citep{2010MNRAS.408..669C} and conduction \citep{2008JCoPh.22710040P}. The model includes subgrid recipes for radiative cooling \citep{2009MNRAS.393...99W}, star formation \citep{2004ApJ...609..667S,2008MNRAS.383.1210S}, stellar evolution and chemical enrichment \citep{2009MNRAS.399..574W}, black hole growth \citep{2005MNRAS.364.1105S,2015MNRAS.454.1038R}, feedback from supernovae \citep{2012MNRAS.426..140D} and active galactic nuclei \citep{2009MNRAS.398...53B}, and reionization \citep{2001cghr.confE..64H,2009MNRAS.399..574W}. We use the flagship Ref-L100N1504 \citep[see][for specifications]{2015MNRAS.446..521S} simulation of a $(100\,\mathrm{cMpc})^3$ volume\footnote{We use $\mathrm{cMpc}$ to denote comoving megaparsecs.} with gas particles of initial mass $1.8\times 10^{6}\,\mathrm{M}_{\odot}$. This simulation adopts the fiducial cosmological parameters of \citet{2014A&A...571A..16P}.

Halos are identified using the `friends-of-friends' algorithm with a linking length equal to $0.2$ times the mean inter-particle spacing \citep{1985ApJ...292..371D}. Gravitationally bound sub-structures (`subhaloes') are subsequently identified using the \textsc{subfind} algorithm \citep{2001MNRAS.328..726S,2009MNRAS.399..497D}. Subhaloes are connected through time to their progenitors and descendents in a merger tree constructed with the \textsc{D-trees} algorithm \citep{2014MNRAS.440.2115J,2017MNRAS.464.1659Q}. We make use of the EAGLE SQL database \citep{2016A&C....15...72M} to query the (sub)halo catalogues and merger trees.

\subsection{Satellites and quenching timescales}\label{subsec:trees}

We construct our sample of rich galaxy groups and galaxy clusters by selecting haloes in the last ($z=0$) snapshot of the simulation with virial\footnote{We adopt the convention that the overdensity defining the virial mass and radius is $200$ times the critical density for closure, $\rho_\mathrm{crit}=3H_0^2/8\pi G$. We also refer to work adopting other definitions, including $360$ times the mean matter density, $\rho_\mathrm{m}=\Omega_\mathrm{m}\rho_\mathrm{crit}$. We distinguish these with subscripts $200\mathrm{c}$ and $360\mathrm{m}$, respectively.} masses $M_\mathrm{200c} > 10^{13}\,\mathrm{M}_{\odot}$. This yields a sample of $167$ host systems, the most massive of which has $M_\mathrm{200c}=10^{14.6}\,\mathrm{M}_{\odot}$ We then select lower-mass galaxies that are within $3.3 r_\mathrm{200c}$ of these host systems and label them `satellites'. $3.3r_\mathrm{200c}$, equivalent to $\approx 2.5r_\mathrm{360b}$ at $z=0$, is about the maximum distance that any satellite galaxies reach at the apocentre of their first orbital passage \citep{2004A&A...414..445M}. Our initial selection includes $4208$ satellites with $9\leq\log_{10}(M_{\star}/\mathrm{M}_{\odot})<11$.

\begin{figure*}
  \includegraphics[width=\textwidth]{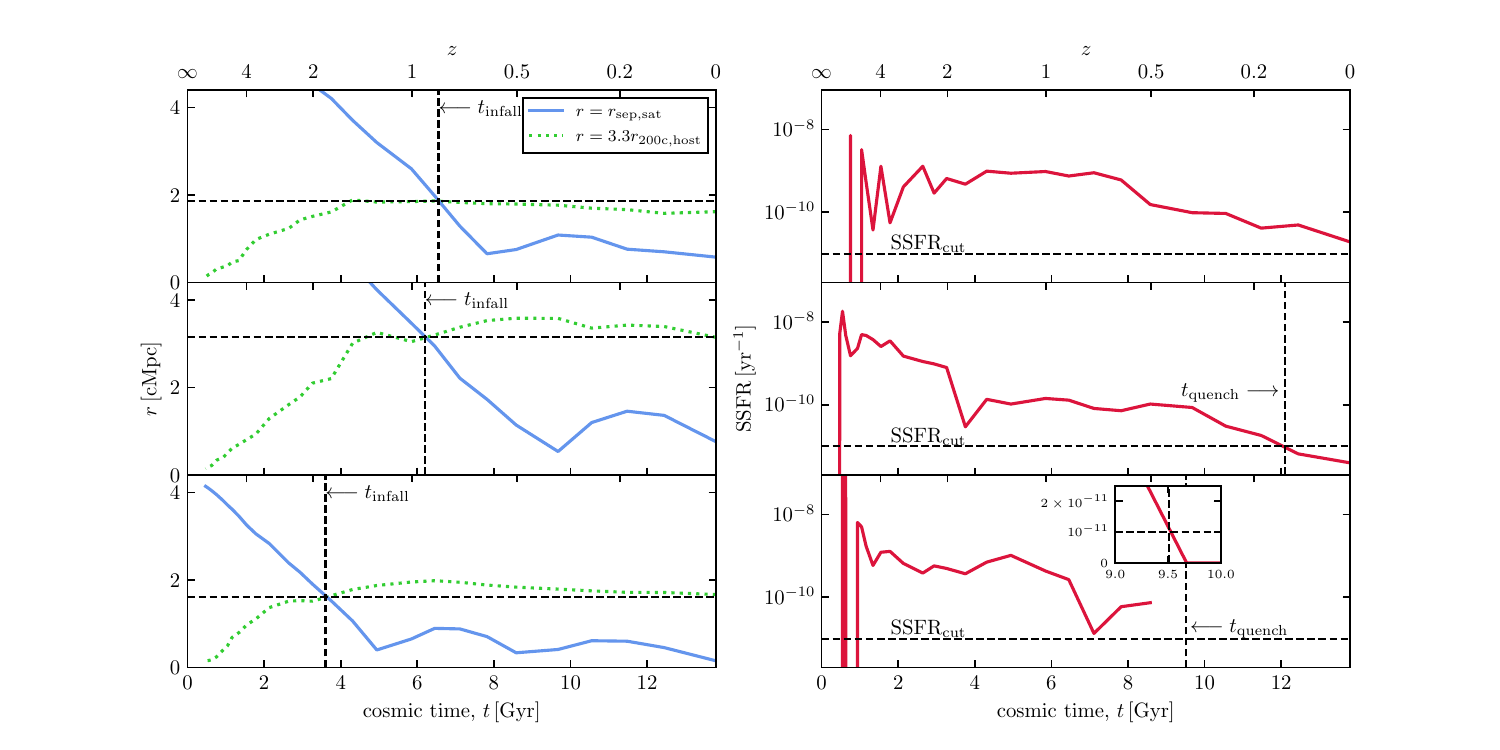}
  \caption{Orbital and star formation histories of three satellite galaxies from our sample. \textit{Left panel:} radial separation between the host and satellite (blue solid line) and $3.3r_{200}$ of the host, the radius defining `infall' in our work (green dotted line), as a function of time. \textit{Right panel:} $\mathrm{SSFR}$ as a function of time. The inset panel shows the same curve as in the lower panel but with linear scaling on both axes. The infall time $t_\mathrm{infall}$ is defined as the first inward crossing of $3.3r_\mathrm{200c}$, and the quenching time $t_\mathrm{quench}$ as the last time when $\mathrm{SSFR}>10^{-11}\,\mathrm{yr}^{-1}$. The satellite shown in the top panels is star forming at $z=0$, so $t_\mathrm{quench}$ is not defined. We linearly interpolate the star formation histories between simulation snapshots; the inset (linear-scaled axes) in the last panel illustrates this for a satellite that sees its $\mathrm{SSFR}$ drop from $>10^{-11}\,\mathrm{yr}^{-1}$ to zero in the space of one simulation output. For quiescent galaxies at $z=0$, we define the quenching timescale $\tau_\mathrm{quench}$ as the difference between $t_\mathrm{quench}$ and $t_\mathrm{infall}$.}
  \label{fig:orbit-and-sfh}
\end{figure*}

We next extract the branches corresponding to the main progenitors \citep[see][sec.~2.2.2, for a definition]{2017MNRAS.464.1659Q} of each host and satellite from the merger trees and use their relative positions to track the orbit of the satellite around the host through time. We label the first (and usually unique) time, linearly interpolated between snapshots, when the satellite approaches within $3.3r_\mathrm{200c}$ of its $z=0$ host system as the `infall time', $t_\mathrm{infall}$, of the satellite. Orbits of three example satellites drawn from our sample are shown in Fig.~\ref{fig:orbit-and-sfh} (left panels). We also track the star formation histories of the satellites along their main progenitor branches -- examples for the same galaxies are shown in the right panels of Fig.~\ref{fig:orbit-and-sfh}. We define the `quenching time', $t_\mathrm{quench}$, of satellites as the last time (linearly interpolated between snapshots) when their specific star formation rate is $\mathrm{SSFR}>10^{-11}\,\mathrm{yr}^{-1}$. This threshold is chosen for simplicity and ease of interpretation -- we have checked that accounting for a mass- and/or redshift-dependent threshold does not qualitatively impact our conclusions (see Sec.~\ref{subsec:caveats}). For galaxies that are quiescent at $z=0$, we define a `quenching timescale` $\tau_\mathrm{quench}=t_\mathrm{quench} - t_\mathrm{infall}$. For satellites that are star forming at $z=0$ (such as the example shown in the upper panels of Fig.~\ref{fig:orbit-and-sfh}) the quenching time and quenching timescale are not defined. We remove satellites with $\tau_\mathrm{quench}<0$, from our sample. The remaining satellite galaxies are those where the environmental influence of the final ($z=0$) host could have had some influence in quenching them, if they are quenched by $z=0$.

We further restrict our sample to satellites with lookback times to infall between $6$ and $9\,\mathrm{Gyr}$. This provides a reasonably large sample in terms of numbers while being a narrow enough interval in time that the satellites are approximately coeval and can reasonably be shifted in time to align their infall times and `stacked' to follow their average properties. Selecting galaxies that fell in at least $6\,\mathrm{Gyr}$ ago also means that we can track the satellites for (at least) $6\,\mathrm{Gyr}$ after infall without any satellites dropping out of the sample. Finally, we remove a small number of galaxies from our sample: (i) cases where it is clear that \textsc{subfind} has misassigned particles between subhaloes in a friends-of-friends group (these manifest with more than $5$ times the mean gas fraction at infall time at fixed stellar mass); (ii) any stripped of more than $90$~per~cent of their stellar mass between infall and $z=0$; and (iii) any that have a gas mass of more than $3$ times their gas mass at infall at a later time (usually these are major mergers). Our final sample includes about $1500$ satellites, of which about $1100$ are quenched at $z=0$.

This selection of satellites is biased in various ways with respect to the general population of satellites, but has the advantage that the average evolution of satellites in the sample is easy to define and to interpret. We defer discussion of the implications of our biased selection to Sec.~\ref{sec:discussion}.

We also construct a control sample of non-satellite galaxies. For each satellite in the sample, we randomly select one non-satellite with a stellar mass at the satellite's infall time within $5$~per~cent of that of the satellite. Non-satellites are restricted to have a subgroup identifier of $>0$ (i.e. times when they are not the most massive subhalo within their friends-of-friends group) during no more than one snapshot after the infall time of the satellite that they are matched to.

\subsection{Particle tracking} \label{sec:particles}

The merger trees tabulate halo and subhalo identifiers for each subhalo at each snapshot time. Together these provide a unique identifier for each subhalo, and the same identifiers are attached as labels to particles in the simulation snapshots. We use these identifiers to extract the list of gas particles identified as bound to each satellite in our sample at its infall time. We do likewise for galaxies in our control sample of non-satellites. Each gas particle in addition has a unique particle identifier that we use to track the extracted particles at each later snapshot. The identifier is preserved even if the gas particle is transformed into a star particle in a star formation event. In the very rare instances where a gas particle is consumed by a supermassive black hole particle, it simply drops out of our list of tracked particles. We track the histories of about $30$ million particles in total.

For each snapshot in the simulation, we record the status of the tracked particles including whether they are still gas particles or have become stars, their coordinates, velocities, halo and subhalo identifiers, temperatures and star formation rates. The last two properties are undefined for particles at times that they are stars. In our analysis below we distinguish three particle categories: star-forming gas (`SF gas') for particles with non-zero star formation rates, non-star-forming gas (`nSF gas') and `stars'. In the EAGLE model, gas particles are assigned a non-zero star formation rate when they exceed a metallicity-dependent density threshold \citep[see][sec.~4.3, for details]{2015MNRAS.446..521S} such that particles with non-zero star formation rates are those with densities and temperatures consistent with being the gas that represents the interstellar medium in the model. We also distinguish between `bound' and `unbound' particles. Bound particles share the halo and subhalo identifier of the descendent of the satellite that they fell in with and are therefore still gravitationally bound to it. Unbound particles have presumably been stripped, ejected by feedback processes, or otherwise lost. Unfortunately accurately labelling particles with the reason that they have been stripped (ram-pressure, tides, viscosity, etc.), or as ejected by feedback, remains remains extremely challenging. For example, most particles `ejected by feedback' in the EAGLE model are not directly heated by feedback events (these are easily labelled) but entrained in the resulting wind (these are very difficult to reliably label); see \citet{2020MNRAS.494.3971M} for a discussion.

In addition to the particles selected as being bound to satellites at their infall times, we also separately record any gas particles not in this set that become bound to satellites in our sample after their infall times. Our analysis below focuses on the fate of the gas that was bound at infall, but tracking this additional gas offers the opportunity to assess how much gas accretion occurs after infall.

\subsection{Analytic galaxy evolution model}\label{subsec:analytic}

To help guide our interpretation of the tracked gas particles selected from satellite galaxies (Sec.~\ref{sec:particles}), we use an analytic galaxy evolution model. We describe it briefly here, with further details included in Appendix~\ref{app:model}. Our model is a modified version of that defined in \citetalias[][eqs.~(1) \& (2)]{2022MNRAS.511.2948M}. Their model is calibrated to provide an accurate (but much simplified) description of galaxy evolution in the EAGLE simulations and is therefore ideally suited to our purposes. Our model begins with gas in two `reservoirs', corresponding to the star-forming and non-star-forming gas bound to a satellite at its infall time in our particle tracking scheme. We identify these two reservoirs with the ISM and CGM, respectively, in the nomenclature of \citetalias{2022MNRAS.511.2948M}. There are two additional mass reservoirs, initially empty, corresponding to gas ejected from the galaxy (notionally beyond the virial radius), and stars. The model describes mass exchanges between the four reservoirs through a set of coupled linear differential equations with coefficients describing the `efficiencies' of relevant physical processes such as gas cooling, star formation and energetic stellar feedback.

Two key features of our model are that (i) it is formulated so that we can track the evolution of gas that was initially in the ISM (i.e. initially star-forming gas) and gas that was initially in the CGM (initially non-star-forming gas) separately, and (ii) we can track the time evolution of gas labelled as bound to a galaxy at a chosen time independent of any freshly accreted gas. We check that the model captures the main features of the evolution of gas in our control sample of central galaxies (see Sec.~\ref{subsec:trees}), then modify the model to apply to satellites with a single change: re-accretion of ejected gas is switched off. We will interpret the output of the model for satellites of different masses below, in Sec.~\ref{subsec:mixing}.

When using our analytic model to make predictions, we assume an initial time of $6.5\,\mathrm{Gyr}$ (after the Big Bang) -- the median infall time of our sample of galaxies whose particles we track (Sec.~\ref{sec:particles}). We make time integration steps using the forward Euler method with a time step of $10\,\mathrm{Myr}$ to evolve the model for $6\,\mathrm{Gyr}$ (the same amount of time for which we track particles) and have verified that the absolute numerical error in the solution for the mass in the various reservoirs relative to the total mass in the model is always less than 1~per~cent (usually much less).

\section{Results}\label{sec:results}

We present our main results pertaining to the quenching timescales of satellite galaxies in the EAGLE simulations (Sec.~\ref{subsec:results-tauq}) and their detailed interpretation through tracking simulation particles (Sec.~\ref{subsec:mixing}), analytic modelling (Sec.~\ref{subsec:interp-model}), and the dependence of the quenching timescale on satellite properties (Sec.~\ref{subsec:results-props}). We also include some additional insights gleaned from our particle tracking method into the fate of stripped stars and gas (Sec.~\ref{subsec:results-icl}).

\subsection{Quenching timescales}\label{subsec:results-tauq}

In Section~\ref{sec:introduction} we introduced the dependence of quenching timescale on the stellar mass of satellite galaxies. We begin by measuring the quenching timescale as a function of satellite stellar mass for our sample of EAGLE satellites in groups and clusters -- the resulting trend is shown with the red cross symbols in Fig.~\ref{fig:obs-qtd}. There is significant scatter (more than $\pm 1\,\mathrm{Gyr}$) in the quenching time at any fixed satellite stellar mass, but the `turnover' in the trend emerges clearly\footnote{The vertical error bars show the interquartile scatter in the distribution. The uncertainty in the median is too small to visualise on the figure: the existence of a `peak' in the quenching timescale near $10^{10}\,\mathrm{M}_{\odot}$ for EAGLE satellite galaxies is not uncertain.} -- satellites with $M_{\star}\sim 10^{10}\,\mathrm{M}_{\odot}$ have the longest quenching timescales.

The `turnover' from increasing quenching timescale with increasing stellar mass for low stellar mass satellite galaxies to decreasing quenching timescale with increasing stellar mass for more massive satellites could hint at a transition between two dominant mechanisms that regulate star formation in satellites (as discussed in Sec.~\ref{sec:introduction}, but we will see below that this is not the best interpretation of the physics at play in EAGLE. Throughout the rest of this work we will build up a picture where the environmental contribution to quenching EAGLE satellite galaxies is driven primarily by `starvation' in the absence of fresh gas accretion compounded by losses through supernova and AGN feedback. In lower stellar mass satellites gas stripping also plays a secondary role (mainly ram pressure stripping, with some contribution from tides), but this only modulates the trends set by feedback.

The quenching timescale is inherently difficult to calibrate absolutely since neither infall nor the cessation of star formation are associated with unambiguous, discrete events. Indeed, the observationally-based inferences shown with greyscale markers/patches in Fig.~\ref{fig:obs-qtd} define both infall and quenching in various ways, which is the leading origin of discrepancies between different studies \citep[see e.g.][sec.~6.2]{2021MNRAS.501.5073O}. Furthermore, the EAGLE simulations do not model the relevant physics with perfect fidelity (further discussed in Sec.~\ref{subsec:caveats}). As an example, `cold' gas discs in EAGLE are too geometrically thick and kinematically hot due to approximations used in the modelling of the ISM. This makes the ISM of EAGLE satellite galaxies artificially more susceptible to ram pressure stripping, for instance. Because of these challenges we focus our analysis on relative trends and on the physical origin of the `turnover' from a positive slope to a negative slope in the scaling of quenching timescale with stellar mass in the simulations.

\subsection{Mass evolution of satellites}\label{subsec:mixing}

\begin{figure*}
  \includegraphics[width=\textwidth]{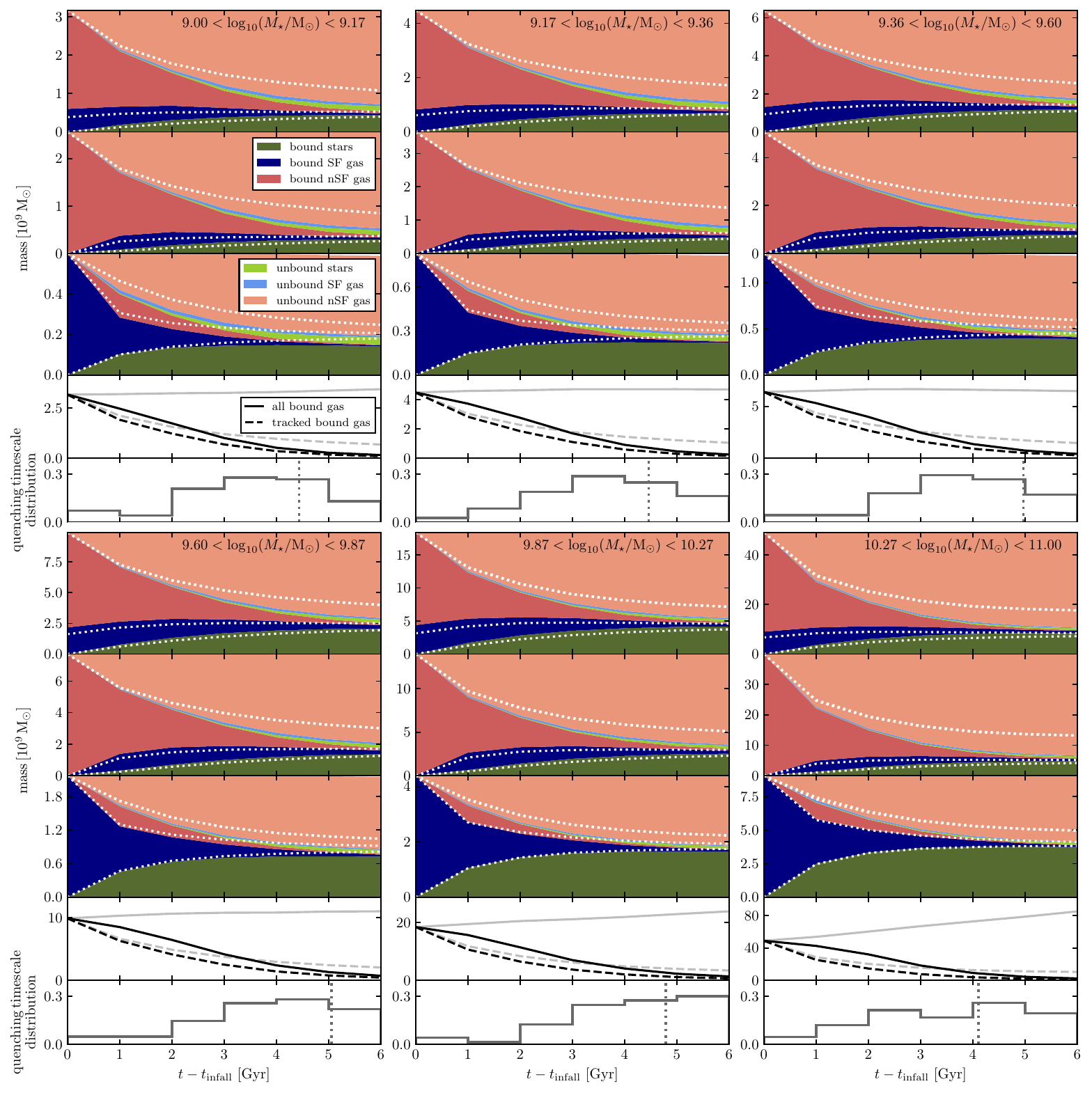}
  \caption{Fate of the gas bound to EAGLE satellite galaxies at their infall times. Each group of $5$ panels is similar and corresponds to one of our stellar mass bins, as labelled. In the upper panel of each group we show the average composition as a function of time of the gas that was bound at the infall time of satellites in that bin. Initially all of the gas is gravitationally bound to the satellite and is either star forming (dark blue) or non-star-forming (dark red). Over time gas may form stars that may be bound (dark green) or not (light green) to the satellite. Unbound (stripped or ejected) gas can be star forming (light blue) or non-star-forming (light red). The second and third panels in each group are similar but decompose the gas into that which was initially non-star-forming or star forming, respectively. The dotted white lines overlaying these panels show the delineation between the same components of gas selected at the same time for our mass-matched sample of central galaxies (normalised to have the same total mass). The fourth panel in each group shows the total bound gas mass as a function of time (solid lines) compared to the bound gas mass of tracked particles (dashed line) -- the black lines are for satellites in our sample, while gray lines are for the mass-matched central galaxies (normalized to the same initial gas mass). The fifth panel in each group shows the distribution of quenching timescales (or equivalently quenching times since $\tau_\mathrm{quench}=t_\mathrm{quench}-t_\mathrm{infall}=0$ is the infall time) of satellites, $\mathrm{d}N(\tau_\mathrm{quench})/N\mathrm{d}t$. The vertical dotted lines mark the quenching timescales calculated from the SFRs of the stacked average of galaxies in each stellar mass bin.}
  \label{fig:mixing-with-central}
\end{figure*}

We begin by visualising the fate of the gas that was bound to EAGLE satellite galaxies at their infall time. Fig.~\ref{fig:mixing-with-central} shows this for each of our $6$ stellar mass bins, spaced such that the galaxy count in each bin is equal. The upper panel in each group shows the evolution of the entire tracked gas mass, averaged over all galaxies in that stellar mass bin. Initially ($t_\mathrm{infall}$, the infall time) all gas is gravitationally bound (darker colour shades) and either star forming (dark blue) or non-star-forming (dark orange). Over time the star-forming gas is converted to gravitationally bound stars (dark green). The bulk of the gas eventually ends up gravitationally unbound (lighter colour shades), predominantly non-star-forming (light orange) with trace amounts of unbound star-forming gas (light blue). The origin of the unbound non-star-forming gas is for the moment ambiguous -- it could in principle be either stripped or ejected -- but we will discriminate between these cases below. The unbound star-forming gas is suggestive of direct stripping of the ISM, although from this figure alone we cannot exclude star-forming gas being entrained in outflows, or gas that is lost from the satellite and later becomes star forming in a different galaxy. There is also a small amount of unbound stars; some of these must be formed from the unbound star-forming gas, but some could also be formed in the satellite and subsequently stripped.

The next two panels in each set separate the gas into that which was non-star-forming at infall time (second panel) and star forming at infall time (third panel). These reveal additional detail of the physics at play. For example, the initially non-star-forming gas can be seen to cool and become star forming (a wedge of dark blue -- star-forming gas -- appears in the second panel, followed by a wedge of dark green -- stars). The rapid depletion of the bound star-forming gas in the third panel resulting in stars and unbound non-star-forming gas is suggestive of star formation feedback-driven outflows: the ratio of stars and unbound non-star-forming gas reached once the star-forming gas is depleted can be loosely interpreted as a stellar wind mass-loading factor ($\eta\approx M_\mathrm{gas,\,unbound}(t_\mathrm{final})/M_\star(t_\mathrm{final})\approx 2$ in the lowest stellar mass bin, decreasing to $\eta\approx1$ in the highest stellar mass bin).

\begin{figure}
  \includegraphics[width=\columnwidth]{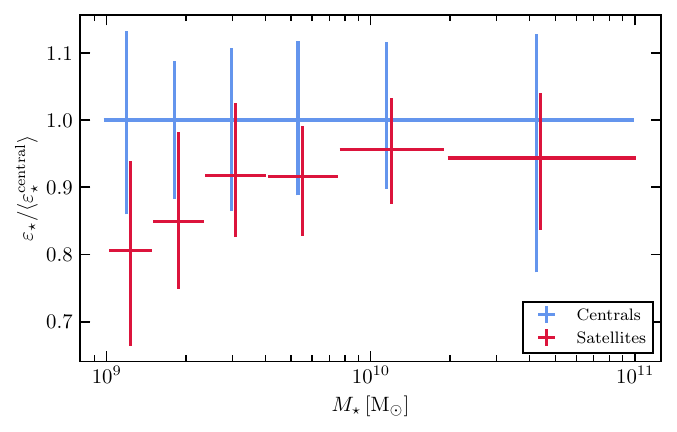}
  \caption{Fraction of the star-forming gas mass at infall converted to stars by $z=0$, $\varepsilon_{\star}$, for satellites in our sample normalized by the average in stellar mass bins of the same quantity for tracked particles in our mass-matched sample of central galaxies, $\langle\varepsilon_{\star}^\mathrm{central}\rangle$ (red crosses). The vertical length of the markers shows the interquartile scatter within the stellar mass bin shown by the horizontal length. The same measurement is shown for the matched sample of central galaxies (blue crosses) to illustrate the scatter in that population -- since the central galaxies are normalized using their own mean $\varepsilon_{\star}/\langle\varepsilon_{\star}^\mathrm{central}\rangle=1$ by definition in this case.}
  \label{fig:sfe}
\end{figure}

Overlaid on these three panels are dotted white curves delineating the boundaries between the same particle classifications for central galaxies in our control sample. These central galaxies have more gas on average than satellites at their infall time, so we have normalized the curves to match the total gas mass in each panel. The trends are broadly very similar -- this is remarkable, as we will discuss further below -- but we begin by highlighting a few differences. First, the fraction of star-forming gas with respect to total gas mass in satellite galaxies is slightly higher at $t_\mathrm{infall}$ than in the control-matched central galaxies (this is actually because they have less non-star-forming gas, rather than more star-forming gas). Central galaxies also retain significantly more bound non-star-forming gas over time than their satellite counterparts. It is tempting to attribute this to stripping of the circumgalactic gas in satellites, but we will see below that this is probably not the dominant driver of this difference. Central galaxies also have no unbound star-forming gas or unbound stars, suggesting that the unbound star-forming gas in the satellites is probably not entrained in winds, but rather reflects direct stripping of the ISM. Finally, the highest stellar mass satellites and central galaxies see their initially bound star-forming gas depleted on a remarkably similar timescale (lower right set of panels, third panel, dark blue wedge), while for lower mass satellites this gas is depleted slightly faster and consequently forms less stars as a fraction of the initial star-forming gas mass. We visualise this trend in Fig.~\ref{fig:sfe} and will interpret it further below.

The reason that the overall similarity between the fate of gas in central and satellite galaxies is remarkable is revealed by the fourth panel in each set in Fig.~\ref{fig:mixing-with-central}. The black curves show the gravitationally bound gas mass in the satellite galaxies at each time including either all gas particles (including any that were not bound at $t_\mathrm{infall}$; solid curve), or considering only the gas particles that were bound at $t_\mathrm{infall}$ (dashed curve). This shows that satellites in our sample accrete a relatively small amount of additional gas after infall, while their overall bound gas mass (solid curve) is dropping precipitously. In the matched sample of central galaxies, on the other hand, the initially bound gas is rapidly becoming unbound (dashed grey curve), but the overall gas supply is being continuously replenished (solid grey curve) such that in the lowest stellar mass bin the total gas mass is approximately constant while in the highest stellar mass bin it grows significantly over the $6\,\mathrm{Gyr}$ that we track the galaxies.

The rapid decay of the tracked bound gas mass in the central galaxies is suggestive of strong winds from stellar or AGN feedback (plus the mass that leaves the gas supply by being converted to stars or accreted by a black hole); presumably central galaxies are not being strongly stripped. Interestingly, this proceeds at a very similar rate to the bound gas mass loss in satellite galaxies, providing a first strong hint that feedback is the main driver of gas loss in the satellites, too. We were surprised at finding such close similarities in the fates of the gas particles that we tracked in central and satellite galaxies: in central galaxies, the large amount of additional accreted gas is also cooling, becoming star forming, feeding an AGN, and therefore triggering energetic feedback. It turns out that separating the galaxies into two components -- the bound gas at $t_\mathrm{infall}$, and everything else -- and considering the evolution of both components `in isolation' works surprisingly well. In mathematical terms, describing the galaxies by a set of coupled linear ordinary differential equations and using the separable property of such equations to focus on a subset of the galaxies is a perhaps surprisingly good approximation\footnote{We had initially hoped that the approximation would be quantitatively good enough to carry out a differential comparison of the tracked bound gas in central and satellite galaxies to isolate the role of stripping. Unfortunately feedback affects all gas in a galaxy indiscriminately, whether we are tracking the particles or not, which precluded this approach. The approximation still holds qualitatively, and we use this extensively in our interpretation in this work.} that we will explore further below with the help of the model outlined in Sec.~\ref{subsec:analytic}.

The last panel in each set in Fig.~\ref{fig:mixing-with-central} shows the distribution of quenching timescales for galaxies in that stellar mass bin -- these are the same distributions represented by the red markers in Fig.~\ref{fig:obs-qtd}. The vertical dotted lines in these panels show the quenching timescale inferred from the stellar mass growth curves in the top panels of each group. The time derivate of these curves corresponds to the star formation rate of the average galaxy in each stellar mass bin, from which we calculate a specific star formation rate and quenching timescale. These quenching timescales differ somewhat from those shown in Fig.~\ref{fig:obs-qtd}: the former are the median quenching times of satellites that have quenched by $z=0$, while the latter are the quenching timescale of the average satellite galaxy, where the average is over all satellites regardless of whether they have quenched by $z=0$ or not. They do, however, maintain the qualitative feature of a maximum in the quenching timescale -- now at $M_\star\sim10^{9.7}\,\mathrm{M}_\odot$ (c.f. $\sim10^{10.0}$ in Fig.~\ref{fig:obs-qtd}). We therefore postulate that understanding the evolution of the average (or `stacked') satellite galaxy is enough to understand the overall evolutionary trends in the population that drive the scaling of the quenching timescale as a function of stellar mass.

\subsection{Interpretation through analytic modelling}\label{subsec:interp-model}

\begin{figure}
  \includegraphics[width=\columnwidth]{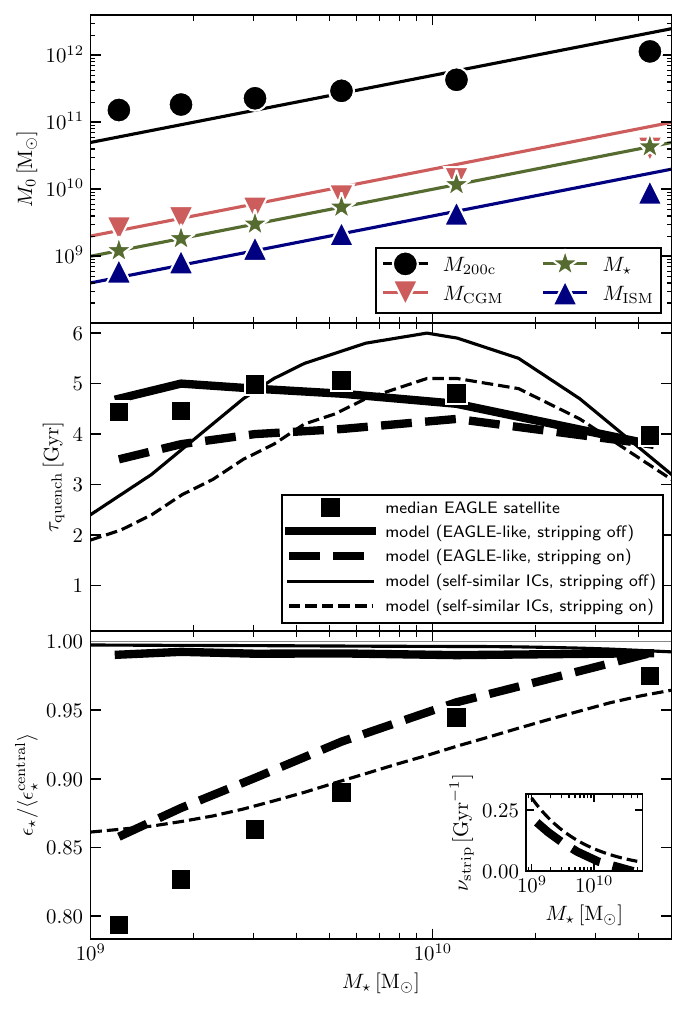}
  \caption{Summary of the results of our analytic model (definition in Sec.~\ref{subsec:analytic}). \textit{Upper panel:} Initial mass in the model galaxies including stellar and halo mass at infall ($M_\star$ and $M_\mathrm{200c}$, respectively) and star-forming ($M_\mathrm{ISM}$) and non-star-forming ($M_\mathrm{CGM}$) gas masses at the same time. The markers show the initialisation values for the `fiducial' models representative of the EAGLE galaxies in our six stellar mass bins. The solid lines show the scaling of the masses for our simplified setup with constant stellar-to-halo mass ratios and gas-to-stellar mass fractions (`self-similar initial conditions (ICs)'; see Sec.~\ref{subsec:interp-model} for details). \textit{Middle panel:} Quenching timescale as a function of stellar mass at infall for our `fiducial' (heavy lines) and self-similar ICs (thin lines) models with our simple stripping prescription switched off (solid lines) or on (dashed lines). A comparable measurement of the quenching timescale derived from our tracking of gas particles in EAGLE satellite galaxies is shown with the square markers. \textit{Lower panel:} Fraction of the star-forming gas mass at infall converted to stars by $z=0$ normalized by the corresponding values for central galaxies; the square markers repeat the values shown with red crosses in Fig.~\ref{fig:sfe}. The solid and dashed lines are as in the middle panel. The inset panel shows the value of our $\nu_\mathrm{strip}$ parameter that provides a simple description of gas stripping in satellites as a function of stellar mass -- this stripping effect is needed to qualitatively reproduce the trend in $\epsilon_\star/\langle\epsilon_\star^\mathrm{central}\rangle$.}
  \label{fig:model-trends}
\end{figure}

The analytic galaxy formation model outlined in Sec.~\ref{subsec:analytic} guides our interpretation of the trends highlighted in Sec.~\ref{subsec:mixing}. As a first step, we demonstrate that the model qualitatively captures the trends shown in Fig.~\ref{fig:mixing-with-central} for satellites and central galaxies. For each of our six stellar mass bins, we initialize a model satellite galaxy that initially has the median stellar mass ($M_\star$), mean star-forming gas mass ($M_\mathrm{ISM}$), mean non-star-forming gas mass ($M_\mathrm{CGM}$) and median total mass ($M_\mathrm{200c}$) of satellites in that bin\footnote{We use the mean for the gas masses because these are directly analogous to the gas masses for the averaged satellites shown in Fig.~\ref{fig:mixing-with-central}. The (initial) stellar and halo masses are instead indicative of what is typical in a given bin, so for these we prefer the median value.}. The initial masses are shown in the upper panel of Fig.~\ref{fig:model-trends} with the six sets of markers. We also initialize model central galaxies according to the properties of central galaxies in our control sample. Integrating these model galaxies forward in time through $6\,\mathrm{Gyr}$ of evolution reproduces many of the broad trends seen in Fig.~\ref{fig:mixing-with-central} including that central galaxies retain more non-star-forming gas for longer than their satellite counterparts. The comparison is discussed further in Appendix~\ref{app:model}.

The model also qualitatively reproduces the scaling of the quenching timescale with stellar mass. The quenching timescale measured from the star formation rate calculated from stars formed from initially bound gas of the average satellite galaxy in each stellar mass bin (i.e. the slope of the boundary of the dark green regions in Fig.~\ref{fig:mixing-with-central} -- we convert this to a specific star formation rate and interpolate to find the first time when $\mathrm{SSFR}<10^{-11}\,\mathrm{yr}^{-1}$) is plotted with the square markers in the middle panel of Fig.~\ref{fig:model-trends}. This differs slightly from the quenching timescales plotted in Fig.~\ref{fig:obs-qtd} which we recall showed the median quenching timescales of satellites quenched by $z=0$. The quenching timescales in Fig.~\ref{fig:model-trends} provide a more fair comparison to the output of the model. There is still a maximum in the quenching timescale around $M_\mathrm{star}=5\times10^{9}\,\mathrm{M}_\odot$. The quenching timescales calculated in the same way from the model galaxies are plotted with the heavy solid lines in the middle panel of Fig.~\ref{fig:model-trends}. The model captures the quenching timescales within a couple of hundred megayears and also predicts a maximum in the quenching timescale, although at a somewhat lower stellar mass ($\sim2\times10^{9}\,\mathrm{M}_\odot$).

This maximum in the quenching timescale in the model may at first glance appear uncertain -- since it occurs in the second-to-lowest stellar mass bin and the slope is rather shallow, does the trend actually turn over? Or is it just due to random `noise', perhaps in the initial model galaxy properties or model coefficients?

We next make a set of `self-similar' model initial conditions (ICs) that will confirm that the model unambiguously predicts a maximum in the quenching timescale, and at the same time unambiguously identify its origin. These initial conditions are self-similar in the sense that the initial ratios in the different mass reservoirs are fixed at $t=0$, for all stellar masses, to to $M_\star/M_\mathrm{200c}=0.02$, $M_\mathrm{ISM}/M_\star=0.4$ and $M_\mathrm{CGM}/M_\star=2$. Fixing these ratios removes any possible influence of the differential properties of the initial model galaxies such as variations in stellar-to-halo mass ratio or gas fractions. These fixed ratios are plotted with (parallel) solid lines in the upper panel of Fig.~\ref{fig:model-trends} and are loosely representative of the average galaxies in each stellar mass bin, as can be seen by comparing the lines to the markers in the figure.

Integrating these model galaxies forward in time and measuring their quenching timescales produces the trend shown with a thin solid line in the middle panel of Fig.~\ref{fig:model-trends}: there is a very clear maximum in the quenching timescale at $M_\star=10^{10}\,\mathrm{M}_\odot$. This maximum can have only one possible origin. Because all of these model galaxies are self-similar in terms of their initial masses and the model equations (Eq.~\ref{eq:model}) have no absolute mass scale, the only possible origin of the peak in the quenching timescale is through the mass-dependence of the model coefficients. Consulting fig.~3 in \citetalias{2022MNRAS.511.2948M}, the parameter describing `halo-scale outflows' ($\eta^\mathrm{halo}$) conspicuously has a minimum at $M_\mathrm{200c}\sim 10^{11.7}$ at all redshifts, corresponding to $M_\star=10^{10.0}$ given our assumed $M_\star/M_\mathrm{200c}=0.02$. The halo mass dependence halo-scale gas outflow rates is therefore the main driver of the behaviour of the quenching timescale as a function of stellar mass.

The $\eta^\mathrm{gal}$ (galaxy-scale outflows) and to a lesser extent $G^\mathrm{gal}_\mathrm{acc}$ (efficiency for galaxy-scale accretion), $G^\mathrm{gal}_\mathrm{ret}$ (efficiency for galaxy-scale return of previously-expelled gas) and $G_\mathrm{SF}$ (star formation efficiency) parameters also have weaker halo-mass dependent features (local extrema or changes in slope) around the same halo mass and therefore also have lesser individual influences. Importantly, the \citetalias{2022MNRAS.511.2948M} model parameters are calibrated against the evolution of only central galaxies from EAGLE. Therefore there is no risk that a parameter labelled `galaxy-scale outflow' ($\eta^\mathrm{gal}$), for instance, is actually adopting values to capture the influence of ram-pressure stripping (for instance) through the calibration process.

The less pronounced maximum in the quenching timescale in our six fiducial models relative to the `self-similar ICs' setup (i.e. comparing the heavy and thin solid lines in the middle panel of Fig.~\ref{fig:model-trends}) is just a consequence of the variation of gas fractions and halo mass as a function of stellar mass. These ratios are not constants as a function of stellar mass in realistic galaxy populations, so their stellar mass dependence also modulates the trend for the quenching timescale as a function of stellar mass.

In summary, the driver of satellite quenching in our analytic model are much the same feedback-driven winds that occur in central galaxies. The only important assumption that we have made in adapting the model for satellite galaxies is that the accretion rate of fresh gas in satellites is zero, and the fourth panel in each set in Fig.~\ref{fig:mixing-with-central} confirms that this assumption is correct -- at least close enough for the purpose of our qualitative argument. The maximum in the quenching timescale is therefore a reflection of the peak in galaxy formation efficiency at halo masses around $10^{12}\,\mathrm{M}_\odot$, with the efficiency of stellar feedback increasing towards lower masses and AGN feedback increasing towards higher masses -- this is the physics in the EAGLE simulations that the scaling of the $\eta^\mathrm{gal}$ and $\eta^\mathrm{halo}$ \citetalias{2022MNRAS.511.2948M} model parameters capture (see their sec.~2.4 for further discussion).

We next consider whether our model allows us to comment on whether satellite-specific processes besides the truncation of accretion, such as ram-pressure or tidal stripping, still have some significant role to play in regulating the quenching timescale of satellites. The clearest feature of satellite galaxy evolution that we have identified that our model fails to even qualitatively reproduce is the integrated efficiency of conversion of ISM gas (that present at infall) into stars. Lower mass satellites convert less of their initial ISM into stars than comparable central galaxies, as introduced in Fig.~\ref{fig:sfe} above. Both of the model variations shown in Fig.~\ref{fig:model-trends} completely fail to reproduce this trend: model satellites have constant a $\epsilon_\star/\langle\epsilon_\star^\mathrm{central}\rangle\approx 1$ as a function of stellar mass\footnote{The value is slightly less than $1$ because of our choice to switch off halo-scale gas recycling ($G^\mathrm{halo}_\mathrm{ret}=0$) for satellites.}, shown by the heavy and thin solid lines in the lower panel of Fig.~\ref{fig:model-trends}. We have also not identified any reasonable modification to our model that could produce this difference between central and satellite galaxies. (Some modifications that we considered while developing the model include: (i) assuming different mixes of pristine and recycled gas by introducing linear combinations of $G^\mathrm{gal}_\mathrm{ret}$ and $G^\mathrm{gal}_\mathrm{acc}$; (ii) assuming a time-dependence for $M_\mathrm{200c}$ instead of holding it constant; (iii) applying a multiplicative scaling to the feedback efficiencies $\eta^\mathrm{gal}$ and $\eta^\mathrm{halo}$; (iv) fixing the redshift to a constant when looking up \citetalias{2022MNRAS.511.2948M} coefficients.) A process that strips gas from satellite galaxies and is more efficient for lower mass satellites, however, offers a well-motivated interpretation of the trend in Fig.~\ref{fig:sfe}. To illustrate this, we add a very simple set of stripping term to the system of equations in our model:
\begin{equation}
\begin{split}
  \dot{M}^\mathrm{ISM,i}_\mathrm{CGM} &= -\nu_\mathrm{strip}M^\mathrm{ISM,i}_\mathrm{CGM}\\
  \dot{M}^\mathrm{ISM,i}_\mathrm{ISM} &= -\nu_\mathrm{strip}M^\mathrm{ISM,i}_\mathrm{ISM}\\
  \dot{M}^\mathrm{CGM,i}_\mathrm{CGM} &= -\nu_\mathrm{strip}M^\mathrm{CGM,i}_\mathrm{CGM}\\
  \dot{M}^\mathrm{CGM,i}_\mathrm{ISM} &= -\nu_\mathrm{strip}M^\mathrm{CGM,i}_\mathrm{ISM}
\end{split}
\end{equation}
The parameter $\nu_\mathrm{strip}$ has dimensions of inverse time. We choose scalings of this parameter that decrease with increasing stellar mass for the two models shown. For the `EAGLE-like' models we choose values in each stellar mass bin -- $\nu_\mathrm{strip}=[0.2, 0.16, 0.12, 0.08, 0.04, 0.00]\,\mathrm{Gyr}^{-1}$ -- that qualitatively reproduce the trend in $\epsilon_\star/\langle\epsilon_\star^\mathrm{central}\rangle$ as shown with the dashed lines in the lower panel of Fig.~\ref{fig:model-trends}. For the `self-similar ICs' models we chose an analytic scaling with a similar shape:
\begin{equation}
\nu_\mathrm{strip}=0.3\exp\left(-2\frac{\log_{10}M_{\star}(t)/M_{\star}(t_\mathrm{infall})}{\log_{10}M_{\star}(t_\mathrm{final})/M_\star(t_\mathrm{infall})}\right).
\end{equation}
The inset panel shows the adopted values of $\nu_\mathrm{strip}$. Adding this stripping term modulates the quenching timescale, but doesn't change its overall qualitative behaviour as a function of stellar mass, as shown by comparing the dashed lines to their corresponding solid lines in the middle panel of Fig.~\ref{fig:model-trends}. We note that in the `self-similar ICs' models (constant halo and gas mass fractions) the quenching timescale changes most at intermediate stellar masses. This has a simple physical explanation: in lower and higher mass satellites, feedback-driven winds quench satellites so efficiently that the stripping barely has time to make its influence felt before feedback in the absence of accretion quenches the galaxies (in addition to our implementation of stripping being intrinsically weak in the most massive satellites).

To summarise our overall interpretation as guided by our analytic modelling efforts: assuming that satellite galaxies evolve much like central galaxies with their halo-scale accretion and recycling shut off, the galaxy- and halo-scale winds driven by energetic feedback are sufficient to explain the quenching timescales of satellite galaxies to leading order. `Environmental processes' beyond the shut off of accretion, such as ram-pressure or tidal stripping, can have a subdominant influence but are not required to explain the main overall trends. We emphasize that these statements apply to population averages; in individual satellites stripping processes may be strong enough to become the dominant quenching mechanism. Indeed, spectacular examples of heavily ram-pressure stripped galaxies are observed \citep[see][for a review]{2022A&ARv..30....3B}, and we find evidence in the simulations for direct stripping of the ISM in at least some of our simulated satellite galaxies (see Sec.~\ref{subsec:results-icl} below).

\subsection{Dependence of quenching timescale on satellite properties}\label{subsec:results-props}

\begin{figure}
  \includegraphics[width=\columnwidth]{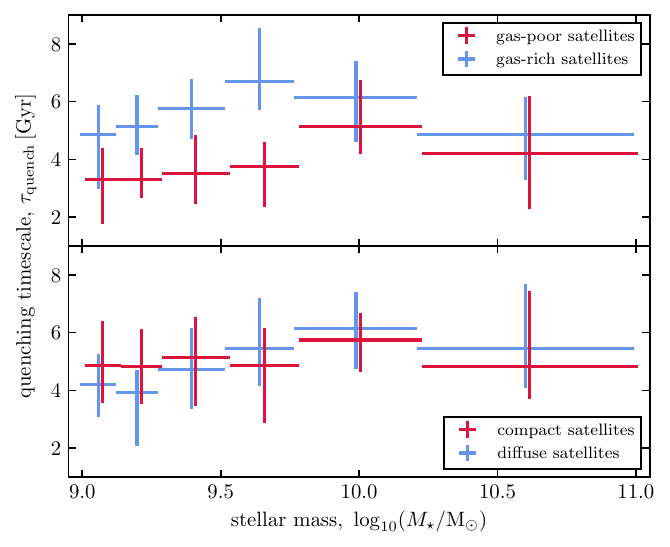}
  \caption{Quenching timescales of EAGLE satellite galaxies as a function of stellar mass. \textit{Upper panel:} Comparison of the quenching timescales of the 35~per~cent most gas-rich (blue crosses) and 35~per~cent most gas-poor (red crosses) satellite galaxies in  each stellar mass bin in our sample. The gas content is measured at the infall time. \textit{Lower panel:} Comparison of the quenching timescales of the 35~per~cent most compact (blue crosses) and 35~per~cent most diffuse (red crosses) satellite galaxies in each stellar mass bin in our sample. Compactness is defined by the radius of the sphere enclosing half of the stellar mass of the satellite.}
  \label{fig:qtd-split}
\end{figure}

We next examine how the scaling of the quenching timescale with stellar mass depends on two intrinsic properties of satellite galaxies -- their gas fractions (with respect to stellar mass) at infall, and their stellar densities. In the upper panel of Fig.~\ref{fig:qtd-split} we plot the trends for the 35~per~cent most gas-rich satellite galaxies in each of six bins of stellar mass alongside those of the 35~per~cent most gas-poor satellites in the same bins. Intuitively, galaxies that become satellites with higher gas fractions are more resilient to quenching at all stellar masses.

\begin{figure}
  \includegraphics[width=\columnwidth]{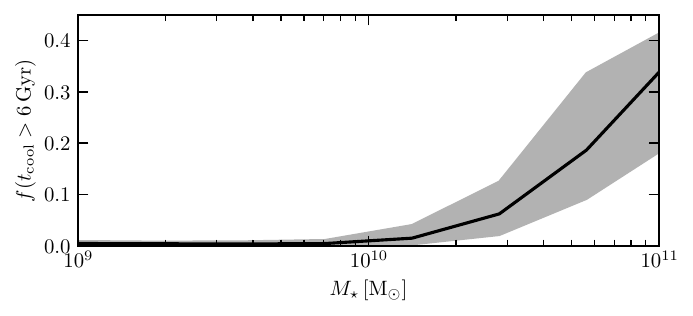}
  \caption{Fraction of non-star-forming gas particles with a cooling timescale longer than $6\,\mathrm{Gyr}$ (the time interval over which we track our sample of satellites galaxies) as a function of stellar mass. Galaxies are binned by stellar mass; the solid line and shaded band illustrate the median and interquartile scatter in each bin.}
  \label{fig:cooling}
\end{figure}

The difference in the quenching timescale between the two groups is small ($\lesssim 1\,\mathrm{Gyr}$) at higher masses but larger at lower and especially intermediate masses ($\sim 3\,\mathrm{Gyr}$). The difference in gas fraction turns out to be mostly driven by the non-star-forming component of the gas: `gas-rich' and `gas-poor' galaxies at fixed stellar mass have the same star-forming gas mass within 50~per~cent or less in all stellar mass bins, while the non-star-forming gas mass is a factor of several larger in the gas-rich group in all bins. In the massive galaxies the time to deplete the star-forming gas reservoir is about the same in gas-rich and gas-poor galaxies because both groups have about the same amount of star-forming gas. The non-star-forming gas plays a lesser role in these galaxies because the cooling times of non-star-forming gas particles are long compared to the quenching timescale (see Fig.~\ref{fig:cooling}), making the non-star-forming gas reservoir mostly inaccessible to star formation. In the lowest and highest mass satellites, the non-star-forming gas can be efficiently removed by feedback processes (stripping also plays a role, especially in lower-mass satellites), quickly making the initially `gas-rich' satellites resemble the `gas-poor' satellites, leaving about equal amounts of more tightly bound star-forming gas. Intermediate-mass satellites can retain their non-star-forming gas for longer, and it can also cool to continue to fuel star formation, maximizing the difference in the quenching time between the `gas-rich' and `gas-poor' groups.

The lower panel of Fig.~\ref{fig:qtd-split} is similar to the upper panel but contrasts the most compact satellite galaxies (defined as the 35~per~cent of the sample within each bin with the smallest stellar half-mass radii) with the most diffuse satellites (35~per~cent with the largest stellar half-mass radii). At low stellar masses diffuse satellites have shorter quenching timescales than their more compact counterparts, while at high stellar masses the trend inverts such that more compact satellites quench more quickly than diffuse satellites of the same stellar mass. More compact satellites are more tightly gravitationally bound, offering an intuitive explanation for the situation at the low stellar mass end: they are more resilient against gas loss through supernova feedback, ram pressure and/or tidal stripping. The reversed trend at the high-mass end is another hint that stripping is not the main driver of the evolution of these objects, instead perhaps compact satellites are more efficient at converting their star-forming gas into stars and so deplete it faster through star formation and the stronger AGN feedback that operates at this mass scale.

\subsection{Contribution to the ICL and BCG}\label{subsec:results-icl}

\begin{figure}
  \includegraphics[width=\columnwidth]{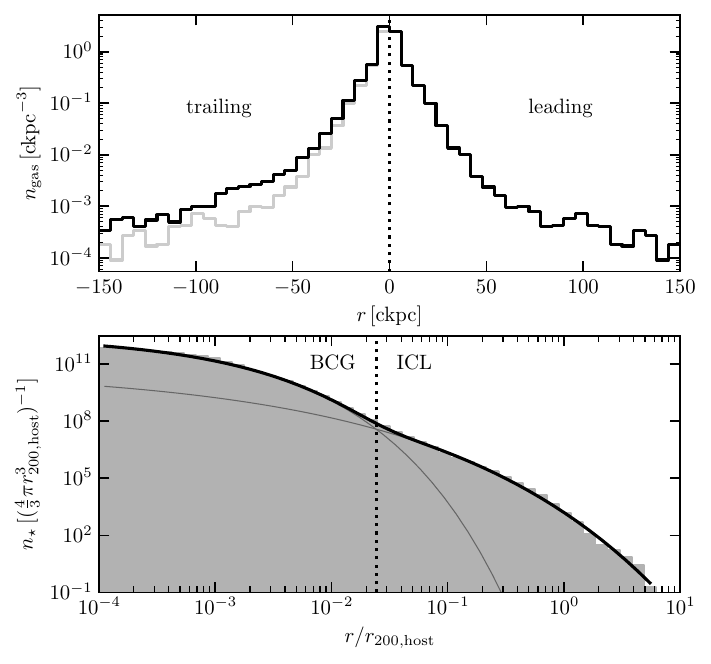}
  \caption{The fate of stripped stars and star-forming gas. \textit{Upper panel:} number density of gas particles that were star forming at the time of the first snapshot when they became unbound from their satellite as a function of their separations from the satellite at $z=0$. A plane (defined at $z=0$) through the centre of the satellite and perpendicular to its velocity vector divides particles into those `leading' the satellite shown with positive separations, and those `trailing' shown with negative separations. The curve for `leading' particles is reflected on the trailing side, shown with the light grey line. There is about twice as much gas on the trailing side. \textit{Lower panel:} number density of star particles either stripped from a satellite in our sample or formed from gas stripped from one of the satellites as a function of radial separation from the host system. The black curve shows a two-component fit to the distribution (the individual components are shown with thinner grey curves), showing that the distribution can be cleanly separated into a compact component that we associate with stars formed from stripped gas that has fallen into the BCG and a diffuse component composed of stars stripped from satellites or formed in star-forming gas tails that contributes to the ICL.}
  \label{fig:stripped-hist}
\end{figure}

The fate of the stars and the star-forming gas stripped from satellite galaxies rounds out our picture of satellite galaxy evolution in EAGLE. In the upper panel of Fig.~\ref{fig:stripped-hist} we examine the spatial distribution at the end of the $6\,\mathrm{Gyr}$ tracking interval of gas particles that were star-forming during the first snapshot after becoming unbound from the satellite galaxy that they fell in with\footnote{This means that the gas is not necessarily star-forming by the time that it reaches the distance where it contributes to the histogram in the figure.}. We bisect each galaxy with a plane passing through its centre and normal to its velocity vector, dividing the gas into a component `trailing' the satellite and another `leading' it. The number density profile of the leading component is repeated with a paler line on the trailing side of the figure to highlight that there is a factor of $\sim 2-3$ more gas on the trailing side. This is compelling evidence of direct ram pressure stripping of the star-forming ISM: tides act approximately symmetrically. The leading component of star-forming gas (at the time when it was stripped) shows that tides also play a role\footnote{Gas that was trailing but now appears to be leading because a satellite has since turned around on its orbit also makes a small contribution.}, but are subdominant by a factor of a few to ram pressure. The figure focuses on the immediate surroundings of the satellites, but the (stacked) distribution of stripped gas continues similarly out to $\sim 1\,\mathrm{Mpc}$ around the satellites.

Stars not bound to satellites but that formed from gas that was bound to a satellite at its infall time can arise through two channels: either stars formed in the satellite and were subsequently stripped from it, or stars formed from gas after it had been ejected or stripped from the satellite. The two channels turn out to contribute about equal amounts to the total mass of stars not bound to satellites that originate from gas that fell in bound to satellites. The lower panel of Fig.~\ref{fig:stripped-hist} show focuses on the `stripped-then-formed' channel, showing the radial distribution of stars around the centre of the host system. There are clearly two components to the distribution which we highlight by fitting a function with two terms of the form:
\begin{equation}
  \tilde{n}_{\star} \exp{\left(-(r/r_{200,\mathrm{host}}\tilde{r})^{\alpha}\right)}
\end{equation}
where $\tilde{r}$ corresponds to a characteristic size of the component, $\alpha$ `stretches' or `compresses' it horizontally and $\tilde{n}_{\star}$ sets its normalization. The `inner' component is dominant within $\sim2$~per~cent of $r_{200\mathrm{c}}$ of the host (typically tens of kiloparsecs), while the outer component extends well past $r_{200\mathrm{c}}$ (to several megaparsecs). We qualitatively label the inner component `BCG' as its size is comparable to that of the brightest cluster galaxy, while the outer component extends far into the intra-cluster light (`ICL'). The two components highlight the different possible fates of stripped gas that eventually forms stars: some eventually falls into a BCG where some star formation is ongoing, while the rest forms stars much further out around the host. Most star formation in stripped gas happens either in the BGC or the diffuse intra-cluster medium (presumably in the gas tails of satellites), although about 26~per~cent of the stripped-then-formed star particles end up either in another subhalo (presumably having formed there; $17$~per~cent) or in the satellite that the gas initially fell in with (presumably stars re-accreted by the satellite after having formed in a gas tail; $5$~per~cent). An important caveat to these measurements is that they are presumably sensitive to the subgrid model assumed for star formation; they could be particularly sensitive to the choice of a metallicity-dependent number density threshold for star formation \citep[][eq.~2]{2015MNRAS.446..521S} in the EAGLE model.

\section{Discussion}\label{sec:discussion}

We discuss some biases and caveats to our results and their interpretation in Sec.~\ref{subsec:caveats}, and compare with the findings of \citet{2019MNRAS.487.3740W} in Sec.~\ref{subsec:wright}.

\subsection{Biases and caveats}\label{subsec:caveats}

Since our main conclusion is that feedback in the absence of accretion, and not direct stripping of ISM gas, is responsible for setting the quenching timescales in satellites of massive groups and clusters, it is important to consider whether the simulations that we use model feedback and stripping faithfully enough to support this. In the case of stripping, this is straightforward. It is reasonably clear that the EAGLE model, like other models with broadly similar resolution and physics implementations, probably strips satellite galaxies somewhat too efficiently \citep{2017MNRAS.470.4186B}. One main reason for this is that limited resolution and the choice to model the ISM with a temperature floor of $8000\,\mathrm{K}$ mean that gas (and as a consequence stellar) discs are thicker and hotter than in reality, making them less tightly bound and more susceptible to stripping. Assuming that stripping in the simulations should be less effective only strengthens our conclusions.

The case of feedback is less clear cut. That energetic feedback is a necessary ingredient in a successful galaxy formation model is relatively uncontroversial. However, how feedback is best implemented and how the outflows that it generates should scale with galaxy mass and as a function of radius in galaxies remains an open problem. The feedback implementation used in EAGLE model heats a relatively small number of particles in the vicinity of a feedback event (AGN or supernova) to a relatively high temperature. How this compares to contemporary cosmological hydrodynamical simulation models is discussed in detail by \citet[][sec.~5.2]{2020MNRAS.494.3971M}. The picture that emerges is that EAGLE exhibits relatively low gas outflow rates on the scale of galactic discs as the few heated particles quickly escape, but they then entrain a much larger amount of circumgalactic gas with them as they travel outwards, generating relatively high outflow rates on the scale of the virial radius. We could therefore speculate that, if anything, feedback in EAGLE might be too weak on scales corresponding to the ISM, and too strong on halo scales. Consulting Fig.~\ref{fig:mixing-with-central}, both gas reservoirs are contribute about equally to the total star formation rate at all stellar masses (i.e. the dark green wedges in the second and third panels in each set of panels have roughly equal amplitudes). It is therefore unclear in which direction an error in the efficacy of feedback in EAGLE would drive our results -- or, the feedback could be reasonably representative of that operating in real galaxies, as hinted (albeit weakly) in sec.~5.3 of \citet{2020MNRAS.494.3971M}. We therefore leave this question open as the most important caveat to our main conclusion.

Our selection criteria (Sec.~\ref{subsec:trees}) introduce some biases into our sample of satellite galaxies. To summarise: we have selected satellites that fell in between $6$ and $9\,\mathrm{Gyr}$ ago (between $z\sim 1.4$ and $0.6$) and survive as satellites at the present day. The peak in the quenching timescale at $M_\star\sim10^{10}\,\mathrm{M}_\odot$ is most pronounced when we measure it from the quenching timescale of satellites that have actually quenched by $z=0$ (Fig.~\ref{fig:obs-qtd}). The peak remains even if we calculate a `characteristic' quenching timescale based on the stacked star formation rates of all satellites (both quenched and unquenched by $z=0$; Fig.~\ref{fig:model-trends}), although it does shift to slightly lower stellar mass ($M_\star\sim5\times10^9\,\mathrm{M}_\odot$), and the amplitude of the peak drops by $\sim 1\,\mathrm{Gyr}$. These shifts broadly reflect the increasing contribution of star-forming galaxies at lower stellar masses, and the asymmetric scatter in the quenching timescale. The latter means that the average quenching timescale of satellites is not the same as the quenching timescale of the average (stacked) satellite. In particular, a small fraction of galaxies quench much faster than can be accounted for by star formation and feedback -- these are either cases where stripping does dominate the quenching process (without contradicting our interpretation of what is typical for the overall population), or galaxies that have been `pre-processed' in smaller groups or clusters before infall. We have made no attempt to label or separate the latter from the overall satellite population. It would be interesting to investigate the role of pre-processing further in future work, but the relatively long ($\geq 2\,\mathrm{Gyr}$) quenching timescales typical of the vast majority of galaxies in our sample suggests that pre-processing is probably not the strongest factor in setting $\tau_\mathrm{quench}$ in these objects.

We also imposed a selection on host haloes to have masses $10^{13} < M_\mathrm{200c}/\mathrm{M}_\odot<10^{14.6}$ (the upper limit is simply the mass of the most massive cluster in the EAGLE simulation that we used). This means that there is a bias in the satellite-to-host mass ratios accessible as a function of stellar mass. For instance, the most massive satellites cannot be found in hosts $1000$ times more massive than themselves because these are too rare to occur in the limited volume of the simulation. However, all satellites in our sample can explore the same range in host halo mass because the halo masses of the most massive satellites in our sample are similar to or less than the halo masses of the least massive groups selected. We expect these considerations to have little impact on our main conclusion: we argue that the most important environmental process regulating the quenching timescale is the cessation of accretion -- this is essentially independent of the mass of or mass ratio between the satellite galaxy and its host. The other main physics at play (star formation and AGN feedback) are predominantly secular processes and therefore insensitive to the host mass.

Our choice of a mass- and redshift-independent cut in specific star formation rate to define quenched galaxies could be criticized as too simple \citep[][provide a helpful discussion in sec.~5.2.1]{2019MNRAS.484.1702P}. We have checked that our main conclusions are not strongly sensitive to this choice. For example, we tried adopting the stellar mass-dependent threshold of \citet[][eq.~13]{2016MNRAS.463.3083O}. Although this significantly reduces the quenching timescales of lower-mass galaxies, our qualitative interpretation is preserved.

Finally, the redshift dependence of the quenching timescale deserves a brief discussion. A subtle point that is often overlooked is that, depending on methodology, the redshift where a quenching time is measured does not necessarily correspond to the redshift range of the observed galaxies constraining the measurement, especially when the quenching timescale is long. For example, in a study of galaxies observed at $z=0$, the measurement could be sensitive to the quenching timescale of galaxies that are undergoing a transition from star forming to passive at the present day, but a different approach could instead be sensitive to the quenching timescales of all galaxies that have become passive by $z=0$. These are not the same. In this work, we have focused on galaxies with a selection in infall time ($6$ to $9\,\mathrm{Gyr}$ ago) and have mostly focused on what can be inferred from the average (stacked) satellite of a given stellar mass regardless of whether it has quenched by $z=0$. These choices are essentially impossible to replicate in observational studies so our results should not be compared quantitatively with these, but we hope that our conclusions from theoretical arguments can help to inform the qualitative interpretation of observation-based studies of the quenching timescale. With some caution due to these considerations, we would say that our conclusions are most applicable to satellites in $z\sim 0$ massive groups and clusters\footnote{However, many of the satellites in our sample do have quenching times that correspond to redshifts between $z\sim0.5$ and $1$. We note that there is evidence that simulations like EAGLE are too efficient at quenching low-mass satellites at redshift $z\sim 1$, see \citet{2023MNRAS.518.4782K}.}.

\subsection{Comparison with \texorpdfstring{\citet{2019MNRAS.487.3740W}}{Wright et al. (2019)}}\label{subsec:wright}

\citet{2019MNRAS.487.3740W} also use the EAGLE simulations to investigate the quenching timescales of central and satellite galaxies. Where there are similarities between our analysis and theirs the measurements qualitatively agree, but our interpretation differs in some places with theirs, so some discussion is appropriate.

First, we note an important difference in definitions. Whereas we define `quenching timescale' as the time between infall and crossing a threshold in $\mathrm{SSFR}$, they define `quenching timescale' as the time between leaving the star-forming main sequence and becoming quenched (defined in terms of colour or $\mathrm{SSFR}$). Measurements in the two studies are therefore not straightforward to compare quantitatively, but we can comment on apparent differences in qualitative conclusions.

They also find a peak in the median quenching timescale of satellite galaxies as a function of their stellar masses (their fig.~11), and observe that the quenching of central and satellite galaxies at the high stellar mass end seems to proceed similarly. This is consistent with our comparison of tracked gas in satellites compared to gas that was present at the same initial time in central galaxies: its evolution proceeds very similarly (lower right set of panels in Fig.~\ref{fig:mixing-with-central}). The central galaxies accrete a large amount of additional gas, but this presumably enters the hot circumgalactic phase and, due to the long cooling time (Fig.~\ref{fig:cooling}), never reaches the star-forming disc.

They go on to observe that in low-mass galaxies, quenching in satellites seems to proceed differently than in central galaxies (when it occurs). They attribute this to the different environment that satellites find themselves in. They tentatively mention stripping of the ISM in satellites as the mechanism responsible, but this is not strongly motivated. We would revise this conclusion, agreeing that the difference in environment is responsible, but that the shut off of accretion and loss of gas to feedback, rather than stripping processes, are the dominant physics setting the quenching timescale.

Finally, they propose a connection between the maximum in the quenching timescale and the gas fraction in satellites (comparing their figs.~7 \& 10). While we agree that the gas fraction can strongly affect the quenching timescale, our experiment of fixing the gas fraction to a constant as a function of stellar mass (`self-similar ICs models discussed in Sec.~\ref{subsec:interp-model}) convinces us that the existence of a peak in the quenching timescale is more closely connected to the scaling of outflow mass loading ($\eta^\mathrm{gal}$ and $\eta^\mathrm{halo}$) with halo mass than it is to the gas fractions of galaxies.

\section{Summary \& conclusions}\label{sec:conclusions}

We measured the quenching timescales ($\tau_\mathrm{quench}$; time intervals between infall and cessation of star formation) of satellite galaxies in the EAGLE cosmological galaxy formation simulations and found that intermediate-mass satellites ($M_\star\sim10^{10}\,\mathrm{M}_\odot$) have longer quenching timescales than higher- and lower-mass satellites (Sec.~\ref{subsec:results-tauq}). We then tracked the subsequent evolution of gas particles that were bound to simulated satellite galaxies at their infall time. This reveals much of the physics at play in the average satellite galaxy (Sec.~\ref{subsec:mixing}), but it remains very challenging to distinguish gas stripped by an external mechanism such as ram pressure from gas expelled by feedback. We used an analytic galaxy formation model adapted from the semi-analytic model of \citetalias{2022MNRAS.511.2948M} to show that the most compelling explanation for the `peak' in the quenching timescale as a function of stellar mass is the minimum in the efficiency of feedback at the same stellar mass (Sec.~\ref{subsec:interp-model}). Our interpretation is therefore that the quenching timescales of most satellite galaxies in this regime are set by the timescale for feedback to heat/eject the ISM as star formation and AGN accretion proceed in the absence of the accretion of fresh gas.

This interpretation does not preclude gas stripping (in particular direct stripping of the ISM; Sec.~\ref{subsec:results-icl}) from playing a meaningful role in the overall evolution of satellite galaxies, but strongly suggests that for most satellite galaxies stripping contributes only a small correction in the narrower context of setting the quenching timescale. The most heavily stripped galaxies, that may be preferentially selected in searches for `jellyfish' galaxies, may be exceptions to this rule without invalidating it at the level of the galaxy population. Our results reinforce the `overconsumption' scenario put forward by \citet[][see also \citealp{2021PASA...38...35C}]{2014MNRAS.442L.105M}.

Our interpretation implies that measuring quenching timescales as the time interval between first pericentric passage (or infall) and the global cessation of active star formation may not be the best approach to understand the role of environment in galaxy quenching. Instead, such metrics may be mostly sensitive to the secular processes -- star formation and feedback in the absence of accretion -- that we argue have the strongest role in setting this timescale. This comes with the caveat that the absence of accretion is environmentally-driven. Other metrics, especially those based on spatially-resolved quantities, are probably better suited to investigating the role of other environmental processes like tidal or ram-pressure stripping \citep[e.g.][]{2019ApJ...872...50L}.

Our approach to following the evolution of simulated satellite galaxies by tracking their gas particles through time is promising \citep[see also][]{2020MNRAS.494.3971M,2020MNRAS.497.4495M,2022MNRAS.511.2600M}, and we have not exhausted its full potential in this work. In particular, much more could be done to exploit the spatially-resolved nature of the galaxies. The limitations of the EAGLE simulations (Sec.~\ref{subsec:caveats}) mean that investigating whether stripping might play a more important role in the outer reaches of the ISM while feedback is dominant in the centres of satellites \citep[see][for a discussion]{2021PASA...38...35C} would probably be asking too much of these simulations, but newer models (e.g. \textsc{colibre}, Schaye et al. in prep; Chaikin et al. in prep) with improved treatment of the ISM make this an interesting avenue for future work. Formulating a complete description of satellite galaxy evolution across mass scales and redshift remains a formidable challenge thanks to the numerous physical processes at play.

\section*{Acknowledgements}

KAO thanks R.~Wright, L.~Cortese and D.~Obreschkow for very helpful discussions, and ICRAR at UWA, where the main conclusions of this work were substantially revised, for its hospitality. The authors also thank P.~Mitchell for sharing code and insight. We thank the referee for a constructive review. MEC's contributions were funded through a Durham University Reimagine Research Initiative summer studentship. KAO acknowledges support by the Royal Society through a Dorothy Hodgkin Fellowship (DHF/R1/231105), and by the European Research Council (ERC) through an Advanced Investigator grant to C.~S.~Frenk, DMIDAS (GA~786910). KAO and MAWV acknowledge support by the Netherlands Foundation for Scientific Research (NWO) through VICI grant 016.130.338 to M.~Verheijen. This work used the DiRAC@Durham facility managed by the Institute for Computational Cosmology on behalf of the STFC DiRAC HPC Facility (www.dirac.ac.uk). The equipment was funded by BEIS capital funding via STFC capital grants ST/K00042X/1, ST/P002293/1, ST/R002371/1 and ST/S002502/1, Durham University and STFC operations grant ST/R000832/1. DiRAC is part of the National e-Infrastructure. This work has made use of NASA's Astrophysics Data System Bibliographic Services.

For the purpose of open access, the author has applied a Creative Commons Attribution (CC BY) licence to any Author Accepted Manuscript version arising from this submission.

Author contributions: AIVZ carried out the analysis of the EAGLE simulations, created the figures (except Fig.~5 by KAO) and led their interpretation in the context of his BSc and MSc dissertations at the University of Groningen. MEC implemented the analytic model and led the interpretation of its results. KAO designed and supervised the project, collated the results and drafted this article based on the dissertations. MAWV contributed to project design and supervision.

\section*{Data availability}

The EAGLE simulations used in this work are publicly released including both (sub)halo catalogues \& merger trees \citep{2016A&C....15...72M} and snapshots of the particle distribution \citep{2017arXiv170609899T}. Further details are available on the web page\footnote{\url{https://icc.dur.ac.uk/Eagle/database.php}} describing the data release.

\bibliographystyle{mnras}
\bibliography{paper}

\appendix

\onecolumn
\section{Analytic model definition}\label{app:model}

In this appendix we describe the analytic galaxy evolution model introduced in Sec.~\ref{subsec:analytic}. We adopt the notation of \citetalias{2022MNRAS.511.2948M} (eq.~(1)), with some modifications as described below.

Because the model of \citetalias{2022MNRAS.511.2948M} is a system of linear ordinary differential equations, formally it is straightforward to separate it into parts that sum to the `full' model, although in practice some care is needed to enforce physical consistency, such as ensuring non-negative mass in all of the reservoirs at all times. This enables the first modification that we make to the original model: we remove the cosmological accretion `source' term by setting $f^\mathrm{halo}_\mathrm{acc}=0$. Any fresh gas entering the system is considered a separate component of the model that we do not follow hereafter. This mimics our selection of gas particles bound to a satellite at infall and our choice to track the fate of this gas and not any additional gas accreted later.

The second modification that we make is to simplify the treatment of `pristine' and `ejected' gas. \citetalias{2022MNRAS.511.2948M} track separately gas that has been in the ISM of any galaxy at any previous time and that which has not. In their model, gas that has been in the ISM in the past is assumed to be enriched with metals and cools more efficiently from the CGM into the ISM. We assume this higher efficiency $G^\mathrm{gal}_\mathrm{ret}$ for gas that is initially in the ISM reservoir (and later enters the CGM and cools), while we assume the lower efficiency $G^\mathrm{gal}_\mathrm{acc}$ for gas initially in the CGM (i.e. assuming that it is unenriched and remains so even if it cools into the ISM and later returns to the CGM). This choice obviates the need for terms involving $M^\mathrm{gal}_\mathrm{ej}$ (effectively dropping their eq.~(2)), which instead become part of the $M_\mathrm{CGM}$ reservoir, and consequently the fraction $F^\mathrm{pr}_\mathrm{CGM}=1$. This simplification through approximation has only a small influence on the time evolution of the model while significantly reducing its complexity.

The third modification that we make is to track separately the gas that is initially in the CGM and the gas that is initially in the ISM. Because the equations are linear each reservoir simply separates into two sub-reservoirs, one for gas initially in the ISM $(M^\mathrm{ISM,i}_\mathrm{CGM}, M^\mathrm{ISM,i}_\mathrm{ISM}, M^\mathrm{ISM,i}_\mathrm{ej}, M^\mathrm{ISM,i}_\star)$ and one for gas initially in the CGM $(M^\mathrm{CGM,i}_\mathrm{CGM}, M^\mathrm{CGM,i}_\mathrm{ISM}, M^\mathrm{CGM,i}_\mathrm{ej}, M^\mathrm{CGM,i}_\star)$, that sum to the total. No mass is ever exchanged between the two sets of sub-reservoirs. This mimics our ability to separately track gas particles in a satellite that were star-forming at its infall time and those that were non-star-forming at its infall time. For conciseness we denote the fraction of gas in some of the sub-reservoirs as follows: $F^\mathrm{ISM,i}_\mathrm{ISM}=M^\mathrm{ISM,i}_\mathrm{ISM}/\left(M^\mathrm{ISM,i}_\mathrm{ISM}+M^\mathrm{CGM,i}_\mathrm{ISM}\right)$; $F^\mathrm{CGM,i}_\mathrm{ISM}=\left(1-F^\mathrm{ISM,i}_\mathrm{ISM}\right)$; $F^\mathrm{ISM,i}_\mathrm{CGM}=M^\mathrm{ISM,i}_\mathrm{CGM}/\left(M^\mathrm{ISM,i}_\mathrm{CGM}+M^\mathrm{CGM,i}_\mathrm{CGM}\right)$; $F^\mathrm{CGM,i}_\mathrm{CGM}=\left(1-F^\mathrm{ISM,i}_\mathrm{CGM}\right)$.

With this separation into components, the feedback terms in the model require careful treatment. When stars form, the model specifies mass flow rates from the ISM to the CGM (galaxy-scale winds with rate $\eta^\mathrm{gal}\frac{G_\mathrm{SF}}{t}M_\mathrm{ISM}$), and from the CGM to the ejected reservoir (halo-scale winds with rate $\eta^\mathrm{halo}\frac{G_\mathrm{SF}}{t}M_\mathrm{ISM}$). Both are proportional to the star formation rate $\frac{G_\mathrm{SF}}{t}M_\mathrm{ISM}$. A problem is apparent: the gas that starts in the ISM immediately forms stars, leading to a non-zero mass flow rate from the CGM to the ejected reservoir, but there is no gas in the CGM yet that was initially in the ISM. The problem is obvious at this initial time, but remains throughout the time integration of the model. We resolve this by calculating the total flow rates (for gas in the two sets of `sub-reservoirs' combined) and then divide the flows between the two sets of sub-reservoirs according to the fraction of mass in the sub-reservoir from which the flow is sourced relative to the total. For example, if star formation triggers ISM outflows towards the CGM at a rate of $1\,\mathrm{M}_\odot\,\mathrm{yr}^{-1}$ and there is currently $10^{10}\,\mathrm{M}_\odot$ in the ISM of which $\frac{1}{4}$ was initially in the ISM and $\frac{3}{4}$ was initially in the CGM, the flow rate between the ISM and CGM sub-reservoirs for gas initially in the ISM is $0.25\,\mathrm{M}_\odot\,\mathrm{yr}^{-1}$, and the flow rate between the ISM and CGM sub-reservoirs for gas initially in the CGM is $0.75\,\mathrm{M}_\odot\,\mathrm{yr}^{-1}$ (recovering the total flow rate of $1\,\mathrm{M}_\odot\,\mathrm{yr}^{-1}$). Conceptually, feedback energy impacts all of the gas in the relevant reservoir and is simply divided up according to how much gas in that reservoir belongs to each sub-reservoir. This introduces terms that look like mass exchanges between the two sets of sub-reservoirs, but the equations enforce that the total mass in a given set of sub-reservoirs is constant in time.

With all of these changes to the \citetalias{2022MNRAS.511.2948M} model implemented, our modified model is defined by the coupled system of equations:
\begin{equation}
\begin{bmatrix}
\dot{M}^\mathrm{ISM,i}_\mathrm{CGM}\\
\dot{M}^\mathrm{ISM,i}_\mathrm{ISM}\\
\dot{M}^\mathrm{ISM,i}_\mathrm{ej}\\
\dot{M}^\mathrm{ISM,i}_{\star}\\
\dot{M}^\mathrm{CGM,i}_\mathrm{CGM}\\
\dot{M}^\mathrm{CGM,i}_\mathrm{ISM}\\
\dot{M}^\mathrm{CGM,i}_\mathrm{ej}\\
\dot{M}^\mathrm{CGM,i}_{\star}
\end{bmatrix}
=
\begin{bmatrix}
-\frac{G^\mathrm{gal}_\mathrm{ret}}{t} & \left(F^\mathrm{ISM,i}_\mathrm{ISM}\eta^\mathrm{gal}-F^\mathrm{ISM,i}_\mathrm{CGM}\eta^\mathrm{halo}\right)\frac{G_\mathrm{SF}}{t} & \frac{G^\mathrm{halo}_\mathrm{ret}}{t} & 0 & 0 & \left(F^\mathrm{ISM,i}_\mathrm{ISM}\eta^\mathrm{gal}-F^\mathrm{ISM,i}_\mathrm{CGM}\eta^\mathrm{halo}\right)\frac{G_\mathrm{SF}}{t} & 0 & 0 \\
\frac{G^\mathrm{gal}_\mathrm{ret}}{t} & -\left(1-R+F^\mathrm{ISM,i}_\mathrm{ISM}\eta^\mathrm{gal}\right)\frac{G_\mathrm{SF}}{t} & 0 & 0 & 0 & -F^\mathrm{ISM,i}_\mathrm{ISM}\eta^\mathrm{gal}\frac{G_\mathrm{SF}}{t} & 0 & 0 \\
0 & F^\mathrm{ISM,i}_\mathrm{CGM}\eta^\mathrm{halo}\frac{G_\mathrm{SF}}{t} & -\frac{G^\mathrm{halo}_\mathrm{ret}}{t} & 0 & 0 & F^\mathrm{ISM,i}_\mathrm{CGM}\eta^\mathrm{halo}\frac{G_\mathrm{SF}}{t} & 0 & 0 \\
0 & \left(1-R\right)\frac{G_\mathrm{SF}}{t} & 0 & 0 & 0 & 0 & 0 & 0 \\
0 & \left(F^\mathrm{CGM,i}_\mathrm{ISM}\eta^\mathrm{gal}-F^\mathrm{CGM,i}_\mathrm{CGM}\eta^\mathrm{halo}\right)\frac{G_\mathrm{SF}}{t} & 0 & 0 & -\frac{G^\mathrm{gal}_\mathrm{acc}}{t} & \left(F^\mathrm{CGM,i}_\mathrm{ISM}\eta^\mathrm{gal}-F^\mathrm{CGM,i}_\mathrm{CGM}\eta^\mathrm{halo}\right)\frac{G_\mathrm{SF}}{t} & \frac{G^\mathrm{halo}_\mathrm{ret}}{t} & 0 \\
0 & -F^\mathrm{CGM,i}_\mathrm{ISM}\eta^\mathrm{gal}\frac{G_\mathrm{SF}}{t} & 0 & 0 & \frac{G^\mathrm{gal}_\mathrm{acc}}{t} & -\left(1-R+F^\mathrm{CGM,i}_\mathrm{ISM}\eta^\mathrm{gal}\right)\frac{G_\mathrm{SF}}{t} & 0 \\
0 & F^\mathrm{CGM,i}_\mathrm{CGM}\eta^\mathrm{halo}\frac{G_\mathrm{SF}}{t} & 0 & 0 & 0 & F^\mathrm{CGM,i}_\mathrm{CGM}\eta^\mathrm{halo}\frac{G_\mathrm{SF}}{t} & -\frac{G^\mathrm{halo}_\mathrm{ret}}{t} & 0 \\
0 & 0 & 0 & 0 & 0 & \left(1-R\right)\frac{G_\mathrm{SF}}{t} & 0 & 0 \\
\end{bmatrix}
\begin{bmatrix}
M^\mathrm{ISM,i}_\mathrm{CGM}\\
M^\mathrm{ISM,i}_\mathrm{ISM}\\
M^\mathrm{ISM,i}_\mathrm{ej}\\
M^\mathrm{ISM,i}_{\star}\\
M^\mathrm{CGM,i}_\mathrm{CGM}\\
M^\mathrm{CGM,i}_\mathrm{ISM}\\
M^\mathrm{CGM,i}_\mathrm{ej}\\
M^\mathrm{CGM,i}_{\star}
\end{bmatrix}.
\label{eq:model}
\end{equation}
For the efficiencies $(G^\mathrm{gal}_\mathrm{acc},G^\mathrm{gal}_\mathrm{ret},G_\mathrm{SF},G^\mathrm{halo}_\mathrm{ret},\eta^\mathrm{halo},\eta^\mathrm{gal},R)$ we assume the same time- and halo mass-dependent values as \citetalias{2022MNRAS.511.2948M} (i.e. plotted in their fig.~3).

Now that we have defined a model that qualitatively reproduces the evolution of gas in our control sample of EAGLE central galaxies, we make one final change to the model to adapt it to satellite galaxies. For central galaxies, we assumed that any gas ejected to large radii (notionally beyond the virial radius, although no radial scale appears explicitly in the model) can return to the CGM with an efficiency proportional to $G^\mathrm{halo}_\mathrm{ret}$. For satellites, we assume that any gas expelled to such large radii is lost and will never return by setting $G^\mathrm{halo}_\mathrm{ret}=0$.

\begin{figure*}
  \includegraphics[width=\textwidth]{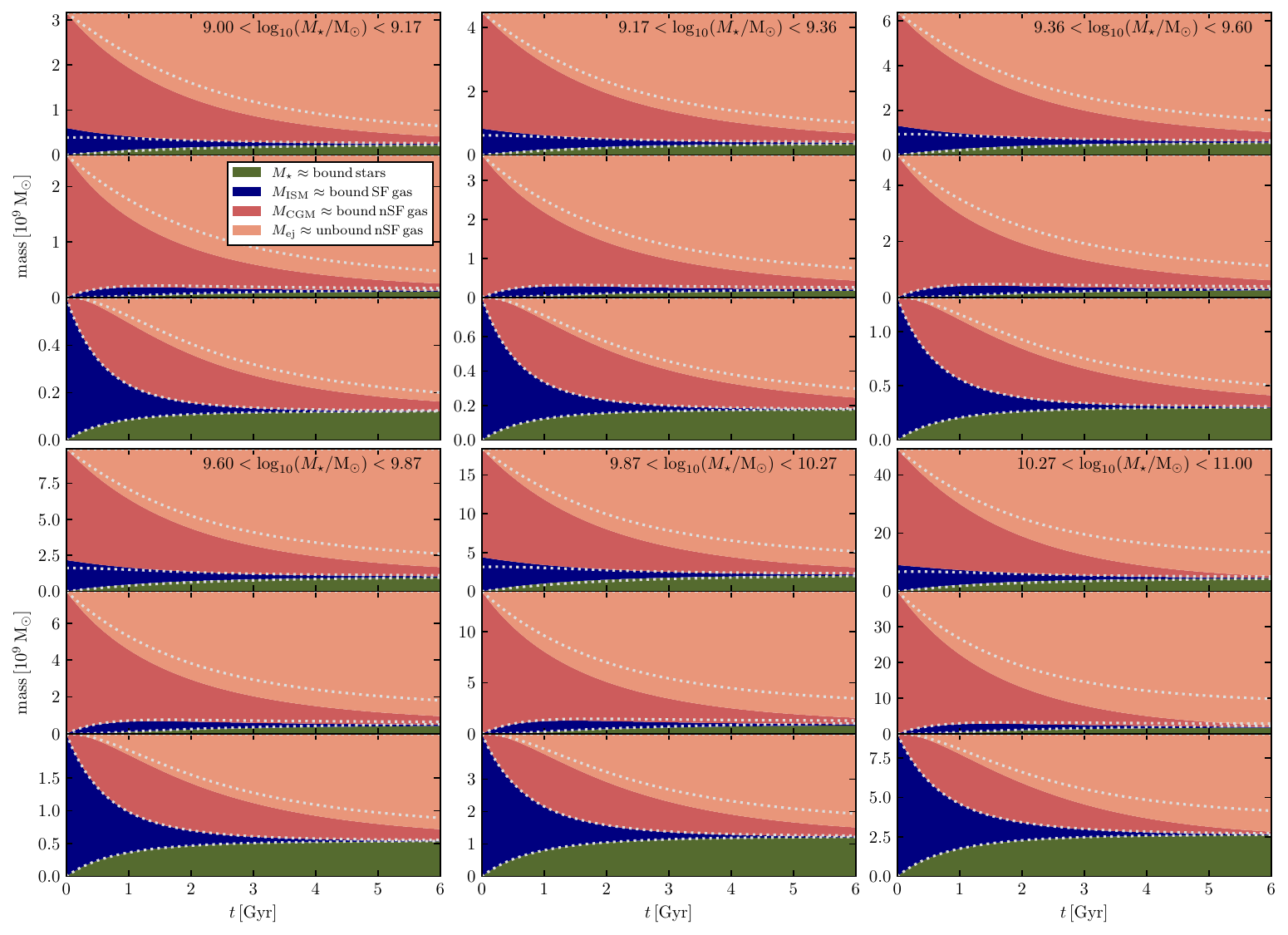}
  \caption{Fate of gas initially bound to satellite galaxies as predicted by our analytic model (Eq.~\ref{eq:model}). Groups of panels are organised as in Fig.~\ref{fig:mixing-with-central} and are comparable to the first three panels in each group in that figure; our analytic model succeeds in capturing the qualitative trends that emerge in the simulated satellite and central galaxies. Each component in the model is conceptually analogous to one of the categories of tracked particles shown in Fig.~\ref{fig:mixing-with-central} as listed in the legend; unbound stars and unbound star-forming gas (light green and blue in Fig.~\ref{fig:mixing-with-central}) have no analogues in the analytic model. In the upper panel in each group we show the evolution of all of the gas initially bound to the satellite galaxy, while the second and third panels decompose this into the portion initially bound to the CGM (superscript $\mathrm{CGM,i}$ in Eq.~\ref{eq:model}) and the ISM (superscript $\mathrm{ISM,i}$), respectively. The model predictions for central galaxies are shown with the dotted white lines marking the boundaries between the same regions as the coloured wedges. The only difference between the calculations for central and satellite galaxies shown in this figure is that for satellites we set $G^\mathrm{halo}_\mathrm{ret}=0$.}
  \label{fig:model-mixing}
\end{figure*}

The evolution of the $6$ models corresponding to our $6$ stellar mass bins (i.e. initial values corresponding to symbols in Fig.~\ref{fig:model-trends}, upper panel) is shown in Fig.~\ref{fig:model-mixing}. Comparing with the similar panels in Fig.~\ref{fig:mixing-with-central}, we find that our model qualitatively reproduces the behaviour of the gas from simulated galaxies that we tracked through time. For example, more massive satellites convert more of their initial ISM mass (dark blue) into stars (green). The main difference between central (dotted white lines) and satellite galaxies (coloured wedges) in the models is that the CGM in satellites is lost faster as it is not replenished by the return of previously ejected gas (controlled by $G^\mathrm{halo}_\mathrm{ret}$). Since our model formulation does not include a treatment of stripping of stars and gas, the unbound stars and star-forming gas components that we track in the simulations (light blue and green in Fig.~\ref{fig:mixing-with-central}) do not occur in the models.

\bsp
\label{lastpage}
\end{document}